\documentclass[aps,prx,reprint,superscriptaddress,floatfix,longbibliography]{revtex4-2}

\usepackage{CJKutf8}
\usepackage[utf8]{inputenc}
\usepackage[T1]{fontenc}
\usepackage[english]{babel}

\usepackage{amsmath,amsfonts,amsthm}
\usepackage{graphicx}
\usepackage{booktabs,makecell}
\usepackage[table]{xcolor}

\usepackage{tcolorbox}
\tcbuselibrary{skins,breakable}
\usepackage[hidelinks]{hyperref}

\newcommand{\Lin}{\mathsf{L}}
\newcommand{\Den}{\mathsf{D}}
\newcommand{\Maps}{\mathsf{T}}

\theoremstyle{plain}
\newtheorem{theorem}{Theorem}[section]
\newtheorem{proposition}[theorem]{Proposition}
\newtheorem{corollary}[theorem]{Corollary}
\newtheorem{lemma}[theorem]{Lemma}
\theoremstyle{definition}
\newtheorem{definition}[theorem]{Definition}

\newcommand{\ourscell}[1]{\cellcolor{black!4}{#1}}
\newcommand{\resultcell}[2]{%
  #1\,{\tiny\textcolor{black!55}{$\pm$ #2}}%
}
\newcommand{\bestresultcell}[2]{\resultcell{\textbf{#1}}{#2}}
\newcommand{\paramcell}[5]{%
  \makecell[c]{#1 / #2 / #3\\#4 / #5}%
}

\newcommand{\tablelayout}[2][3.5pt]{%
  \small
  \setlength{\tabcolsep}{#1}%
  \renewcommand{\arraystretch}{#2}%
}

\newsavebox{\tablecontentbox}
\newcommand{\fittable}[1]{%
  \begingroup
  \sbox{\tablecontentbox}{#1}%
  \ifdim\wd\tablecontentbox>\linewidth
    \sbox{\tablecontentbox}{%
      \resizebox{\linewidth}{!}{\usebox{\tablecontentbox}}%
    }%
  \fi
  \ifdim\dimexpr\ht\tablecontentbox+\dp\tablecontentbox\relax>0.75\textheight
    \resizebox*{!}{0.75\textheight}{\usebox{\tablecontentbox}}%
  \else
    \usebox{\tablecontentbox}%
  \fi
  \endgroup
}

\newcommand{\equalcontribution}{%
  Y. Song and Z. Yang contributed equally to this work and should be considered as co-first authors.}

\graphicspath{{figs/}}

\newcommand{\widefiguregraphics}[1]{%
  \includegraphics[
    width=\linewidth,
    height=0.75\textheight,
    keepaspectratio
  ]{#1}%
}


\begin{document}
\begin{CJK*}{UTF8}{gbsn}
\title{Harnessing Dephasing for Quantum Reservoir Computing with a Minimum Complexity Architecture: Applications to Nonlinear Time-Series Prediction}

\author{Yicheng Song (宋奕成)}
\thanks{\equalcontribution}
\affiliation{Department of Computer Science, Nagoya Institute of Technology, Nagoya 466-8555, Japan}
\email[Contact author: ]{song.yicheng@nitech.ac.jp}

\author{Zhipeng Yang (杨智鹏)}
\thanks{\equalcontribution}
\affiliation{Beijing Academy of Quantum Information Sciences, Beijing 100193, China}
\email[Contact author: ]{yangzp@baqis.ac.cn}

\author{Heng Fan (范桁)}
\affiliation{Beijing Academy of Quantum Information Sciences, Beijing 100193, China}
\affiliation{Institute of Physics, Chinese Academy of Sciences, Beijing 100190, China}
\author{Gouhei Tanaka ({\CJKfamily{min}田中剛平})}
\affiliation{Department of Computer Science, Nagoya Institute of Technology, Nagoya 466-8555, Japan}
\affiliation{International Research Center for Neurointelligence, The University of Tokyo, Tokyo 113-0033, Japan}

\begin{abstract}
Quantum reservoir computing (QRC) exploits quantum dynamics for temporal information processing. Quantum noise affects the reservoir dynamics and can either impair computational performance or serve as a computational resource. However, the role of pure dephasing in Hamiltonian encoding QRC remains unclear. In this work, we propose a minimum complexity QRC model that harnesses pure dephasing in an open uniform XX chain. We show that, under suitable input encoding, pure dephasing can ensure the echo state and fading memory properties even without amplitude damping. Furthermore, in the presence of amplitude damping, the proposed quantum reservoir has distinct stationary states for distinct inputs, and introducing pure dephasing alters these states, thereby affecting the reservoir dynamics. Information processing capacity analysis shows that joint control of the two decoherence mechanisms regulates memory and nonlinearity, providing access to diverse capacity profiles. The proposed QRC achieves promising performance on synthetic benchmarks, while its 6-qubit configuration outperforms all three multi-reservoir echo state network baselines on over half of the real-world prediction tasks. These findings highlight pure dephasing as a computational resource for QRC.
\end{abstract}
\maketitle
\end{CJK*}

\section{Introduction}
Reservoir computing (RC) is a machine learning framework for processing temporal data, including time-series prediction, classification, and anomaly detection \cite{jaeger2001echo,maass2002real,lukovsevivcius2009reservoir}. It consists of a dynamic reservoir that maps inputs into high-dimensional features in an input-history-dependent manner and a trainable readout that maps these features to desired outputs. The reservoir is fixed and only the readout is trained with a simple method such as linear regression and classification. The reservoir is often implemented in software using recurrent neural networks, whereas it can be implemented in hardware based on physical systems that exhibit input-dependent dynamic responses \cite{tanaka2019recent}. Thus, the major advantages of RC are its low training cost compared with deep learning models that rely on backpropagation, and the wide range of options for hardware implementation, which offers significant potential for future AI devices \cite{yan2024emerging}.

Quantum reservoir computing (QRC) is a form of physical RC in which quantum systems, typically multiqubit systems, serve as reservoirs. It exploits quantum dynamics in high-dimensional state spaces to generate input-dependent features, which are extracted through measurements and processed by a readout. The exponentially growing state space of multiqubit systems offers the potential to generate rich features with relatively few qubits. QRC was first proposed by Fujii and Nakajima in 2017, employing a fully connected transverse-field Ising model as the reservoir~\cite{Fujii2017}. At each time step, one qubit is reset to an input-dependent state before unitary evolution, while the remaining qubits retain information about previous inputs. This state-encoding scheme has been widely adopted in subsequent studies, becoming a mainstream approach to QRC~\cite{Nakajima2019,MartinezPena2021}. An alternative approach is Hamiltonian encoding, in which temporal inputs are encoded directly into the generator of continuous-time reservoir dynamics~\cite{sannia2024dissipation}.

Understanding the role of quantum noise and harnessing its effects are important for designing effective quantum reservoirs for near-term devices. Using temporal information processing capacity analysis, Kubota et al. showed that amplitude damping channels can induce the ability to process temporal information in quantum reservoirs~\cite{Kubota2023}. Building upon this insight, Fry et al. further showed that optimizing the hyperparameters of reset noise channels can improve QRC performance on nonlinear benchmarks~\cite{Fry2023}. To implement the abstract noise channels, Ricci et al. realized tunable amplitude damping channels by applying controlled rotations~\cite{Ricci2026}. On the experimental side, Hou et al. implemented QRC with amplitude damping noise on a nuclear magnetic resonance platform~\cite{Hou2026}. In all the aforementioned studies, noise channels are applied after input encoding. However, these studies have not investigated the impact of continuous-time quantum noise induced by interaction with an external environment on the quantum reservoir.

The Hamiltonian-encoding approach provides a scheme in which continuous-time quantum noise can act simultaneously with input driving, allowing such noise to be explored as a computational resource~\cite{sannia2024dissipation}. On the theoretical side, Mart\'{\i}nez-Pe\~na and Ortega provided a framework for analyzing the echo state property (ESP) and the fading memory property (FMP) of finite-dimensional QRC, which provides a theoretical foundation for this approach~\cite{martinezpena2023finite}. Furthermore, Sannia et al. constructed a continuous-dissipation QRC model based on local amplitude damping, established its universal approximation property, and evaluated its performance on benchmarks~\cite{sannia2024dissipation}. In addition, G\"otting et al. analyzed the relationship between amplitude damping rate and linear and nonlinear memory capacities~\cite{gotting2025connection}. As with amplitude damping, pure dephasing is a common decoherence mechanism in practical qubit systems~\cite{onizhuk2025decoherence}, and its effect can be adjusted in some implementations~\cite{krantz2019quantum}. However, the role of pure dephasing in the theoretical properties and computational performance of the Hamiltonian-encoding approach remains unclear.

The dynamics and computational performance of QRC also depend on the underlying physical system, including continuous-variable systems and qubit networks. Continuous-variable systems, including single nonlinear oscillator~\cite{Govia2021}, coupled oscillator pair~\cite{Dudas2023}, and oscillator network~\cite{GarciaBeni2023}, offer infinite-dimensional local Hilbert spaces. In contrast, qubit networks, a mainstream choice in QRC, have two-dimensional local Hilbert spaces. The interqubit coupling range and interaction model shape information propagation and reservoir dynamics. Previous studies have explored qubit network reservoirs with long-range coupling~\cite{Fujii2017, gotting2025connection}, as well as nearest-neighbor coupling~\cite{Li2026}. Besides the coupling range, reservoirs also differ in their interaction models, including Ising~\cite{Franceschetto2026} and XX~\cite{gotting2025connection} models. In most studies, coupling strengths are disordered across the coupled pairs. Given the variety of qubit network interactions, a minimum complexity QRC architecture that can balance structural simplicity and computational performance needs to be studied.

To address these questions, we propose a QRC model that harnesses pure dephasing with a minimum complexity architecture. We adopt an open uniform XX qubit chain as the reservoir and include both local amplitude damping and pure dephasing. Following the Hamiltonian-encoding approach, inputs are encoded into local longitudinal fields through a bipolar input weight matrix. Theoretically, we revisit sufficient conditions for the ESP and the FMP, examine the necessity of the mixing time in Hamiltonian encoding QRC, and analyze the relationship between pure dephasing and these two properties. We also investigate how input and pure dephasing affect the stationary states of the quantum reservoir. Then, we analyze how both decoherence mechanisms and other system hyperparameters affect memory and nonlinearity by evaluating information processing capacity (IPC). Finally, we assess basic computational performance on synthetic benchmark tasks and compare the proposed QRC with multi-reservoir echo state networks (MRESNs) on real-world prediction tasks.

The remainder of this paper is organized as follows. Section~\ref{sec:proposed-qrc} introduces the proposed QRC model. Section~\ref{sec:esp-fmp} revisits sufficient conditions for the ESP and the FMP and examines the necessity of a mixing time and the role of pure dephasing in Hamiltonian encoding QRC. Section~\ref{sec:memory-nonlinearity} investigates how the two decoherence mechanisms and other system hyperparameters affect memory and nonlinearity by IPC. Section~\ref{sec:benchmark-tasks} evaluates the proposed QRC on synthetic benchmark tasks. Section~\ref{sec:prediction-tasks} compares the proposed QRC with MRESNs on real-world prediction tasks. Section~\ref{sec:discussion-conclusion} concludes the paper.

\section{Proposed QRC}
\label{sec:proposed-qrc}
\begin{figure*}[!tp]
  \centering
  \widefiguregraphics{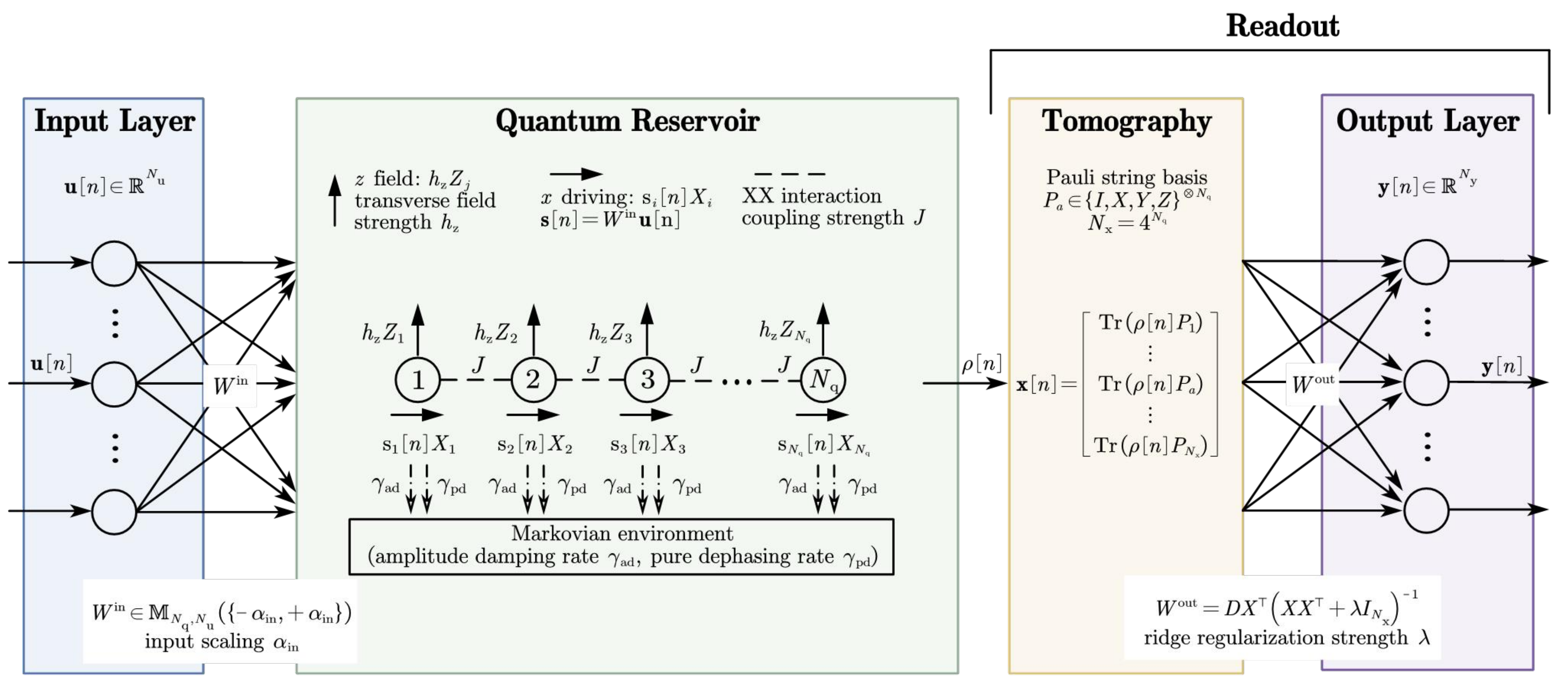}
  \caption{Schematic of the proposed QRC system.}
  \label{fig:qrc_framework}
\end{figure*}

We propose a QRC system in which inputs are encoded by a bipolar input weight matrix into site-dependent, time-dependent local longitudinal fields that drive the reservoir qubits. The quantum reservoir is modeled as an open XX chain in a transverse magnetic field with open boundary conditions and uniform nearest-neighbor couplings. The dynamical evolution of the open quantum reservoir system is governed by a Lindblad master equation incorporating both local amplitude damping and pure dephasing. At each discrete time step, the input-dependent Lindblad evolution updates the reservoir state, which is subsequently mapped to a feature vector through Pauli-basis tomography. Then, a trainable output weight matrix transforms the resulting features into desired outputs. The overall architecture and information processing workflow are illustrated in Fig.~\ref{fig:qrc_framework}, while the input layer, quantum reservoir, and readout are detailed in the following subsections.

The proposed system has three defining features. First, we introduce a bipolar input weight matrix that encodes an input vector into local fields. In contrast to the input schemes adopted in conventional dissipative QRCs~\cite{sannia2024dissipation,martinezpena2023finite}, which are formulated primarily for scalar inputs, the proposed encoding naturally accommodates multidimensional inputs while retaining minimal input layer complexity. Second, inspired by the minimum complexity principle of simple cycle reservoirs~\cite{rodan2011minimum,li2024simplecycle}, we model the quantum reservoir as a uniform XX chain. By dispensing with the disordered couplings and fully connected topology employed in conventional dissipative QRCs~\cite{sannia2024dissipation,gotting2025connection}, our construction provides a structurally simple and deterministic design that may also facilitate physical implementation. Third, we extend conventional amplitude-damping-based QRCs~\cite{sannia2024dissipation,gotting2025connection} by incorporating pure dephasing as a controllable resource. As demonstrated in Secs.~\ref{sec:memory-nonlinearity},~\ref{sec:benchmark-tasks}, and~\ref{sec:prediction-tasks}, both amplitude damping and pure dephasing regulate the reservoir’s memory and nonlinearity, while jointly tuning their rates yields competitive performance across multiple tasks. Moreover, since amplitude damping and pure dephasing commonly coexist in physical implementation~\cite{onizhuk2025decoherence,krantz2019quantum}, their joint inclusion provides a more realistic open system model and strengthens the practical relevance of the proposed system.

\subsection{Input layer}
\label{subsec:qrc-input-layer}
At discrete time step $n$, the input layer encodes an input vector $\mathbf{u}[n]\in\mathbb{R}^{N_{\mathrm{u}}}$ into local driving amplitudes for the quantum reservoir, where $N_{\mathrm{u}}$ denotes the input dimension. We employ a bipolar input weight matrix $W^{\mathrm{in}}\in \mathbb{M}_{N_{\mathrm{q}}\times N_{\mathrm{u}}} (\{-\alpha_{\mathrm{in}},+\alpha_{\mathrm{in}}\})$, where $\mathbb{M}_{m\times n}(S)$ denotes the set of all $m\times n$ matrices with entries in set $S$, $N_{\mathrm{q}}$ is the number of reservoir qubits, and $\alpha_{\mathrm{in}}>0$ is the input scaling parameter. The entries of $W^{\mathrm{in}}$ may be assigned according to a prescribed deterministic rule or sampled independently and uniformly from $\{-\alpha_{\mathrm{in}},+\alpha_{\mathrm{in}}\}$. Unless otherwise stated, we adopt the latter scheme. Once constructed, $W^{\mathrm{in}}$ remains fixed throughout each experiment trial. The encoded input is $\mathbf{s}[n]=W^{\mathrm{in}}\mathbf{u}[n] \in\mathbb{R}^{N_{\mathrm{q}}}$. Consequently, each component $s_i[n]$ specifies the driving amplitude applied to the $i$th reservoir qubit. These amplitudes enter the reservoir Hamiltonian as local longitudinal magnetic fields along the $x$ direction.

\subsection{Quantum reservoir}
\label{subsec:qrc-reservoir}

We consider a quantum reservoir consisting of an open XX chain of $N_{\mathrm{q}}$ qubits in a uniform transverse magnetic field with open boundary conditions and a uniform nearest-neighbor coupling strength. Let $X_i$, $Y_i$, and $Z_i$ denote the Pauli operators that act on the $i$th qubit and as the identity on all other qubits, and let $\sigma_i^-$ denote the corresponding lowering operator. The Hamiltonian of the chain reads
\begin{equation}
  H_{\mathrm{sys}}
  =
  J\sum_{i=1}^{N_{\mathrm{q}}-1}
  \left(
    X_i X_{i+1}+Y_i Y_{i+1}
  \right)
  +
  h_{\mathrm{z}}\sum_{i=1}^{N_{\mathrm{q}}}Z_i,
  \label{eq:qrc-system-hamiltonian}
\end{equation}
where $J\in\mathbb{R}$ is the coupling strength and $h_{\mathrm{z}}\in\mathbb{R}$ is the transverse field strength. During the $n$th input, the encoded amplitudes drive the chain through the input-dependent Hamiltonian $H_{\mathrm{in}}[n]=\sum_{i=1}^{N_{\mathrm{q}}}s_i[n]X_i$. Then, the total Hamiltonian is $H[n]=H_{\mathrm{sys}}+H_{\mathrm{in}}[n]$.

We consider two local dissipative processes acting on each reservoir qubit: amplitude damping and pure dephasing. Accordingly, the reservoir state evolves according to the Lindblad master equation
\begin{equation}
  \begin{aligned}
    \frac{\mathrm{d}\rho(t)}{\mathrm{d}t}
    &=
    L_{\mathbf{u}[n]}\bigl(\rho(t)\bigr)
    \\
    &=
    -i\left[H[n],\rho(t)\right]
    \\
    &\quad+
    \sum_{\mu}
    \left(
      L_\mu\rho(t)L_\mu^\ast
      -
      \frac{1}{2}
      \left\{
        L_\mu^\ast L_\mu,\rho(t)
      \right\}
    \right),
  \end{aligned}
  \label{eq:qrc-lindblad-equation}
\end{equation}
where ${\left(\cdot\right)}^\ast$ denotes the operator adjoint, $L_{\mathbf{u}[n]}$ is the input-dependent Lindblad superoperator of input $\mathbf{u}[n]$, $[A,B]=AB-BA$, and $\{A,B\}=AB+BA$. The jump operators are $\{L_\mu\} =\{L_i^{\mathrm{ad}},L_i^{\mathrm{pd}} \mid i=1,\ldots,N_{\mathrm{q}}\}$, where $L_i^{\mathrm{ad}} =\sqrt{\gamma_{\mathrm{ad}}}\,\sigma_i^-$ describes local amplitude damping with rate $\gamma_{\mathrm{ad}}\geq0$, and $L_i^{\mathrm{pd}} =\sqrt{\gamma_{\mathrm{pd}}}\,Z_i$ describes local pure dephasing with rate $\gamma_{\mathrm{pd}}\geq0$. Let $\rho[n-1]$ denote the initial state for the evolution driven by the $n$th input. Over an  evolution time $\Delta t$, the dynamics generated by $L_{\mathbf{u}[n]}$ gives the discrete reservoir state update
\begin{equation}
  \rho[n]
  =
  \exp\!\left(
    \Delta t\,\,L_{\mathbf{u}[n]}
  \right)
  \bigl(\rho[n-1]\bigr).
  \label{eq:qrc-discrete-update}
\end{equation}

For a fixed number of qubits, the hyperparameters introduced by the input layer and quantum reservoir  are $\alpha_{\mathrm{in}}$, $J$, $h_{\mathrm{z}}$, $\Delta t$, $\gamma_{\mathrm{ad}}$, and $\gamma_{\mathrm{pd}}$. From Eq.~\eqref{eq:qrc-discrete-update}, the reservoir evolution depends on these parameters only through the dimensionless products $\alpha_{\mathrm{in}}\,\Delta t$, $J\,\Delta t$, $h_{\mathrm{z}}\,\Delta t$, $\gamma_{\mathrm{ad}}\,\Delta t$, and $\gamma_{\mathrm{pd}}\,\Delta t$. We therefore choose $\Delta t$ as the unit of time and set $\Delta t=1$ without loss of generality. For notational simplicity, we retain the same symbols for the resulting dimensionless parameters, so that $\alpha_{\mathrm{in}}$, $J$, $h_{\mathrm{z}}$, $\gamma_{\mathrm{ad}}$, and $\gamma_{\mathrm{pd}}$ constitute the five effective hyperparameters of the proposed QRC.

\subsection{Readout}
\label{subsec:qrc-readout}

For the readout, we first perform Pauli tomography on the reservoir state $\rho[n]$. The feature associated with each Pauli string $P_a\in{\left\{I,X,Y,Z\right\}}^{\otimes N_{\mathrm{q}}}$ is defined as its expectation value $x_a[n]=\operatorname{Tr}\!\left(\rho[n]P_a\right)\in\mathbb{R}$, where $\operatorname{Tr}$ denotes the trace. Since the Pauli basis for an $N_{\mathrm{q}}$-qubit system consists of $N_{\mathrm{x}}=4^{N_{\mathrm{q}}}$ elements, the resulting feature vector is $\mathbf x[n]={\left(x_1[n],x_2[n],\ldots,x_{N_{\mathrm{x}}}[n]\right)}\in\mathbb R^{N_{\mathrm{x}}}$. Then, the QRC output is obtained by $\mathbf y[n]=W^{\mathrm{out}}\mathbf x[n]$, where $W^{\mathrm{out}}\in \mathbb R^{N_{\mathrm{y}}\times N_{\mathrm{x}}}$ is a trainable output weight matrix, and $N_{\mathrm{y}}$ denotes the output dimension. The output weight matrix $W^{\mathrm{out}}$ is learned using ridge regression. For a training sequence of length $M$, let $\mathbf d[n]\in\mathbb R^{N_{\mathrm{y}}}$ denote the target output at discrete time step $n$, and define the feature and target matrices as $X=[\mathbf x[1]\ \mathbf x[2]\ \cdots\ \mathbf x[M]] \in\mathbb R^{N_{\mathrm{x}}\times M}$ and $D=[\mathbf d[1]\ \mathbf d[2]\ \cdots\ \mathbf d[M]] \in\mathbb R^{N_{\mathrm{y}}\times M}$, respectively. The output weight matrix is obtained by minimizing the ridge-regression cost function
\begin{equation}
  E_{\mathrm{RR}}(W)
  =\left\|D-WX\right\|_{\mathrm{F}}^2
  +\lambda\left\|W\right\|_{\mathrm{F}}^2,
  \label{eq:qrc-ridge-objective}
\end{equation}
over $W\in\mathbb{M}_{N_{\mathrm{y}}\times N_{\mathrm{x}}}(\mathbb R)$, where $\lambda\geq0$ is the ridge regularization strength and $\|\cdot\|_{\mathrm{F}}$ denotes the Frobenius norm. For $\lambda>0$, the unique minimizer of $E_{\mathrm{RR}}(W)$ is given in closed form by
\begin{equation}
  W^{\mathrm{out}}
  =DX^{\top}
  {\left(XX^{\top}+\lambda I_{N_{\mathrm{x}}}\right)}^{-1},
  \label{eq:qrc-ridge-regression}
\end{equation}
where ${\left(\cdot\right)}^{\top}$ and ${\left(\cdot\right)}^{-1}$ denote the matrix transpose and inverse, respectively, and $I_{m}$ denotes the $m\times m$ identity matrix.

The particular choice of the Pauli-string basis is not essential. Let $\mathcal{B}={\left\{B_\mu\right\}}_{\mu=1}^{N_{\mathrm{x}}}$ and $\mathcal{C}={\left\{C_\mu\right\}}_{\mu=1}^{N_{\mathrm{x}}}$ be two orthonormal self-adjoint operator bases. Since both bases are orthonormal, they are related by an orthogonal matrix $T\in\mathbb{M}_{N_{\mathrm{x}}\times N_{\mathrm{x}}}(\mathbb R)$ such that $C_\mu=\sum_{\nu=1}^{N_{\mathrm{x}}}T_{\mu\nu}B_\nu$. By the linearity of the trace, their feature matrices obey $X_{\mathcal{C}}=TX_{\mathcal{B}}$. For the same target data and ridge regularization strength $\lambda$, Eq.~\eqref{eq:qrc-ridge-regression} then gives $W_{\mathcal{C}}^{\mathrm{out}} =W_{\mathcal{B}}^{\mathrm{out}}T^{\top}$. Consequently, the corresponding outputs satisfy $W_{\mathcal{C}}^{\mathrm{out}}X_{\mathcal{C}} =W_{\mathcal{B}}^{\mathrm{out}}X_{\mathcal{B}}$. Thus, although the feature coordinates and trained output weights depend on the chosen operator basis, the resulting output is invariant. Besides the Pauli basis, another commonly used basis is the generalized Gell-Mann basis, which is used in Ref.~\cite{martinezpena2023finite} to obtain a Bloch-vector state-affine representation of finite-dimensional QRC systems.

\section{Echo state and fading memory properties}
\label{sec:esp-fmp}
Two fundamental properties commonly studied in reservoir computing are the echo state property and the fading memory property. Intuitively, the ESP characterizes the reservoir’s ability to uniformly forget its initial condition as the input history becomes sufficiently long~\cite{yildiz2012revisiting}, whereas the FMP means that inputs that are close in the recent past produce similar outputs, regardless of differences in the distant past~\cite{boyd1985fading}. In theoretical analyses, the ESP is equivalently characterized by the existence and uniqueness of solutions. An RC system satisfying the ESP defines a unique causal and time-invariant filter. The FMP is equivalently characterized by the continuity of this filter and therefore implies the ESP. Both properties are directly related to the universal approximation ability of RC systems. Further theoretical details can be found in Refs.~\cite{grigoryeva2018echo,ortega2026echoes}.

In this section, we proceed in two steps. In Subsec.~\ref{subsec:general-finite-dimensional-qrc}, following Ref.~\cite{martinezpena2023finite}, we review sufficient conditions for the ESP and the FMP of general finite-dimensional QRC systems and the construction methods for systems satisfying these conditions. We also provide an alternative proof of Theorem~\ref{thm:part2-positive-contractive}, which states that strict positivity implies strict contractivity, and prove Proposition~\ref{prop:part2-uniform-primitivity-index}, which establishes the existence of a common exponent such that the power of the auxiliary reservoir map is strictly positive uniformly over the inputs. In Subsec.~\ref{subsec:qrc-quantum-markov-semigroups}, we specialize these results to QRC systems induced by quantum Markov semigroups, a class that includes the QRC proposed in Sec.~\ref{sec:proposed-qrc}. We revisit the mixing time $\Delta t_{\mathrm{mix}}$ introduced in Ref.~\cite{sannia2024dissipation} and show that no such threshold is required. 
Then, we establish three properties of the proposed QRC under suitable input encoding. First, we show that the proposed QRC satisfies the ESP and the FMP even without amplitude damping. Second, we show that distinct inputs lead to distinct stationary states in the presence of amplitude damping. Third, we show that introducing pure dephasing alters the stationary states under amplitude damping alone, thereby affecting the reservoir dynamics. To this end, we prove Proposition~\ref{prop:part2-proposed-reservoir-algebra}, Lemma~\ref{lem:hamiltonian-stationary-states} and Lemma~\ref{lem:ad-pd-stationary-state},  all of which apply more generally to XYZ reservoirs.

\subsection{General finite-dimensional QRC systems}
\label{subsec:general-finite-dimensional-qrc}
Throughout this section, let $\mathcal X$ be a complex Euclidean space. We denote by $\Lin(\mathcal X)$ the set of linear operators on $\mathcal X$, by $\Den(\mathcal X)$ the set of density operators on $\mathcal X$, and by $\Maps(\mathcal X)$ the set of linear maps on $\Lin(\mathcal X)$. Operator inequalities are understood in the Löwner order: for self-adjoint operators $A,B\in\Lin(\mathcal X)$, we write $A\succeq B$ if $A-B$ is positive semidefinite and $A\succ B$ if $A-B$ is positive definite. The trace norm is denoted by $\|\cdot\|_1$. A quantum channel is a completely positive and trace-preserving map on $\Lin(\mathcal X)$.

Following Ref.~\cite{martinezpena2023finite}, a finite-dimensional QRC system is specified by a reservoir map $T:\Lin(\mathcal X)\times\mathbb R^{N_{\mathrm{u}}}\to\Lin(\mathcal X)$ and a readout map $h:\Lin(\mathcal X)\to\mathbb R^{N_{\mathrm{y}}}$. For every input $\mathbf u\in\mathbb R^{N_{\mathrm{u}}}$, the map $T(\cdot,\mathbf u)$ is a quantum channel, while $T$ and $h$ are continuous. We next introduce three channel properties that will be used to establish the ESP and the FMP.

\begin{definition}[Strict contractivity, strict positivity, and primitivity]
Let $\Phi:\Lin(\mathcal X)\to\Lin(\mathcal X)$ be a quantum channel.
\begin{enumerate}
  \item[\textnormal{(i)}] The channel $\Phi$ is \emph{strictly contractive} if there exists a constant $c\in[0,1)$ such that $\|\Phi(\rho)-\Phi(\sigma)\|_1\le c\|\rho-\sigma\|_1$ for all $\rho,\sigma\in\Den(\mathcal X)$.
  \item[\textnormal{(ii)}] The channel $\Phi$ is \emph{strictly positive} if every nonzero positive semidefinite operator is mapped to a positive definite operator, i.e., $A\succeq 0$ and $A\ne0$ imply $\Phi(A)\succ0$ for all $A\in\Lin(\mathcal X)$.
  \item[\textnormal{(iii)}] The channel $\Phi$ is \emph{primitive} if $\Phi^k$ is strictly positive for some $k\in\mathbb N$. The minimum such $k$ is called the \emph{index of primitivity} and is denoted by $w(\Phi)$.
\end{enumerate}
\end{definition}
To verify strict positivity, it suffices to consider density operators. Furthermore, a quantum channel is primitive if and only if it is mixing and admits a faithful fixed state. Consequently, every primitive quantum channel is irreducible \citep{singh2025zeroerror}.

For general RC systems, strict contractivity is a standard sufficient condition for the ESP and the FMP \citep{grigoryeva2018echo}. The same implication holds for QRC systems.

\begin{proposition}[Strict contractivity criterion for ESP and FMP {\citep[Proposition~1]{martinezpena2023finite}}]\label{prop:part2-contractivity-esp-fmp}
Let $D_{N_{\mathrm{u}}}\subset\mathbb R^{N_{\mathrm{u}}}$ be a compact input set, and let $T:\Lin(\mathcal X)\times\mathbb R^{N_{\mathrm{u}}}\to\Lin(\mathcal X)$ be the reservoir map of a QRC system. If $T(\cdot,\mathbf u)$ is a strictly contractive channel for each input $\mathbf u\in D_{N_{\mathrm{u}}}$ with a common contraction constant $c\in[0,1)$, then the QRC with reservoir map $T$ has the ESP and the FMP.
\end{proposition}

In fact, the common contraction constant need not be assumed separately. Indeed, compactness of $D_{N_{\mathrm{u}}}$ and continuity of $T$ imply that pointwise strict contractivity of $T(\cdot,\mathbf u)$ for all $\mathbf u\in D_{N_{\mathrm{u}}}$ yields a common constant $c\in[0,1)$.

The following theorem shows that strict positivity implies strict contractivity. Therefore, we focus on strict positivity as a sufficient condition for a QRC system to possess the ESP and the FMP.

\begin{theorem}[Strict positivity implies strict contractivity]\label{thm:part2-positive-contractive}
Let $\Phi:\Lin(\mathcal X)\to\Lin(\mathcal X)$ be a strictly positive channel. Then there exists $c\in[0,1)$ such that $\|\Phi(\rho)-\Phi(\sigma)\|_1\le c\|\rho-\sigma\|_1$ for all $\rho,\sigma\in\Den(\mathcal X)$.
\end{theorem}

\begin{proof}
A proof can be found in \citep[Theorem~6.3]{rahaman2020new}. See Appendix~\ref{app:strict-positivity-contractivity} for an alternative proof.
\end{proof}

For a single primitive channel $\Phi$, the power $\Phi^{w(\Phi)}$ is strictly positive. For a QRC system whose map $T(\cdot,\mathbf u)$ is primitive for every $\mathbf u$, we seek a common exponent $W(T)$ satisfying $W(T)\ge w(T(\cdot,\mathbf u))$ for all $\mathbf u$. The powered reservoir map ${T(\cdot,\mathbf u)}^{W(T)}$ can then be used to construct a QRC with ESP and FMP.

\begin{proposition}[Uniform exponent for a primitive reservoir map]\label{prop:part2-uniform-primitivity-index}
Let $D_{N_{\mathrm{u}}}\subset\mathbb R^{N_{\mathrm{u}}}$ be a nonempty compact set, and let $T$ be a QRC reservoir map such that $T(\cdot,\mathbf u)$ is primitive for every $\mathbf u\in D_{N_{\mathrm{u}}}$. Then there exists $W(T)\in\mathbb N$ such that ${T(\cdot,\mathbf u)}^{W(T)}$ is strictly positive for every $\mathbf u\in D_{N_{\mathrm{u}}}$.
\end{proposition}

\begin{proof}
See Appendix~\ref{app:uniform-primitivity-index}.
\end{proof}

Therefore, one way to construct a QRC with the ESP and the FMP is to choose an auxiliary primitive reservoir map and define the reservoir map as a power of the auxiliary map.

\begin{corollary}[Construction of reservoir maps with ESP and FMP]\label{cor:part2-general-construction}
Let $D_{N_{\mathrm{u}}}\subset\mathbb R^{N_{\mathrm{u}}}$ be a nonempty compact set, and let $\overline T$ be an auxiliary reservoir map such that $\overline T(\cdot,\mathbf u)$ is primitive for every $\mathbf u\in D_{N_{\mathrm{u}}}$. Let $W(\overline T)$ be the common power as in Proposition~\ref{prop:part2-uniform-primitivity-index}, and define $T(\cdot,\mathbf u):={\overline T(\cdot,\mathbf u)}^{W(\overline T)}$. Then $T(\cdot,\mathbf u)$ is strictly positive for every $\mathbf u\in D_{N_{\mathrm{u}}}$. Consequently, the QRC system with reservoir map $T$ has the ESP and the FMP.
\end{corollary}

\subsection{QRC systems induced by quantum Markov semigroups}
\label{subsec:qrc-quantum-markov-semigroups}
We now consider QRC systems whose reservoir maps are generated by quantum Markov semigroups. Note that the QRC proposed in Sec.~\ref{sec:proposed-qrc} belongs to this class.

\begin{definition}[Quantum Markov semigroup]
Let $\Phi={\{\Phi_t\}}_{t\ge0}$ be a family of linear maps on $\Lin(\mathcal X)$. We say that $\Phi$ is a quantum Markov semigroup (QMS) in the Schrödinger picture if
\begin{enumerate}
  \item[\textnormal{(i)}] $\Phi_0$ is the identity operator on $\Lin(\mathcal X)$;
  \item[\textnormal{(ii)}] $\Phi_{t+s}=\Phi_t\Phi_s$ for all $s,t\ge0$;
  \item[\textnormal{(iii)}]  $\Phi$ is  continuous in the induced trace norm $\|\cdot\|_{1}$, i.e., $\lim_{t\to0^+}\|\Phi_t-\Phi_0\|_{1}=0$;
  \item[\textnormal{(iv)}] $\Phi_t$ is completely positive and trace preserving for every $t\ge0$.
\end{enumerate}
\end{definition}

The corresponding QMS in the Heisenberg picture is defined by the adjoint maps with respect to the trace-duality pairing. More precisely, if ${(\Phi_t)}_{t\ge0}$ is a Schrödinger-picture QMS, then the corresponding Heisenberg-picture semigroup ${(\Phi_t^\ast)}_{t\ge0}$ is defined by $\operatorname{Tr}(\Phi_t(A)B)=\operatorname{Tr}(A\Phi_t^\ast(B))$ for all $A,B\in\Lin(\mathcal X)$. In this correspondence, trace preservation of ${(\Phi_t)}_{t\ge0}$ is equivalent to unitality of ${(\Phi_t^\ast)}_{t\ge0}$, while complete positivity is preserved under taking adjoints.

The generator of a QMS is characterized by the Gorini–Kossakowski–Sudarshan–Lindblad (GKSL) representation.

\begin{theorem}[GKSL representation {\citep[Sec.~4, Theorem~4]{chruscinski2022dynamical}}]\label{thm:part2-gksl-representation}
Let $d=\dim\mathcal X$. The generator $L\in\Maps(\mathcal X)$ of a QMS $\Phi={\{\Phi_t\}}_{t\ge0}$ in the Schrödinger picture has the form
\begin{equation}
  \begin{aligned}
    L(X)
    &=-i[H,X]
    \\
    &\quad+\sum_{j=1}^{N}\left(
      L_j X L_j^\ast
      -\frac{1}{2}\{L_j^\ast L_j,X\}
    \right),
  \end{aligned}
  \label{eq:part2-gksl-representation}
\end{equation}
for $X\in\Lin(\mathcal X)$, where $N\in\mathbb N_0=\{0,1,2,\ldots\}$, $H\in\Lin(\mathcal X)$ is self-adjoint, and $L_j\in\Lin(\mathcal X)$. The operators $H$ and $L_j$ can be chosen traceless, with $N\le d^2-1$. Conversely, if $L\in\Maps(\mathcal X)$ has the form in Eq.~\eqref{eq:part2-gksl-representation} for some self-adjoint $H$ and operators $L_j$, then $\Phi={\{\Phi_t=e^{tL}\}}_{t\ge0}$ is a QMS in the Schrödinger picture.
\end{theorem}

In the Heisenberg picture, the GKSL representation still holds. The generator $L^\ast$ is the Banach adjoint of $L$ given by $L^\ast(X)=i[H,X]+\sum_{j=1}^{N}(L_j^\ast X L_j -\frac{1}{2}\{L_j^\ast L_j,X\})$ for all $X\in\Lin(\mathcal X)$.

We next introduce two closely related properties that will be used to establish the ESP and the FMP for QMS-based QRC systems.

\begin{definition}[Positivity improving and relaxation]
Let $\Phi={\{\Phi_t\}}_{t\ge0}$ be a QMS.
\begin{enumerate}
  \item[\textnormal{(i)}] The QMS $\Phi$ is \emph{positivity improving} if $X\succeq0$ and $X\ne0$ imply $\Phi_t(X)\succ 0$ for every $t>0$.
  \item[\textnormal{(ii)}] The QMS $\Phi$ is \emph{relaxing} if there exists a density operator $\rho$ such that $\lim_{t\to+\infty}\Phi_t(\eta)=\rho$ in trace norm for every density operator $\eta$.
\end{enumerate}
\end{definition}

\begin{theorem}[Equivalence of positivity improving and relaxation {\citep[Theorem~4.2]{fagnola2026irreducibility}}]\label{thm:part2-qms-equivalence}
For QMSs, positivity improving and relaxing with a faithful steady state are equivalent.
\end{theorem}

For a Markovian open quantum system, the dynamics is described by a QMS generated by a Lindbladian $L$. Consider an input-dependent Lindbladian $L_{\mathbf u}$, for example obtained by injecting the input into the Hamiltonian $H_{\mathbf u}$. For a fixed evolution time $\Delta t>0$, we can define a reservoir map $T(X,\mathbf u)=\exp(\Delta t\,\,L_{\mathbf u})(X)$. In Ref.~\citep{sannia2024dissipation}, relaxing QMSs are considered. For each input $\mathbf u$, one can define a mixing time $\Delta t_{\mathrm{mix}}(\mathbf u)$ such that, for every $\Delta t\ge\Delta t_{\mathrm{mix}}(\mathbf u)$, the corresponding reservoir map $T(\cdot,\mathbf u)$ is strictly contractive. By an argument analogous to the proof of Proposition~\ref{prop:part2-uniform-primitivity-index}, compactness of the input domain and continuity of the input-dependent dynamics yield a common mixing time $\Delta t_{\mathrm{mix}}$ that is valid for all admissible inputs. Note that, in many cases, including the setting considered in Ref.~\citep{sannia2024dissipation}, the QMS is relaxing and admits a faithful steady state. By Theorem~\ref{thm:part2-qms-equivalence}, such a QMS is positivity improving. Therefore, $T(\cdot,\mathbf u)$ is strictly positive for every $\Delta t>0$, so no mixing time threshold is required. This is consistent with the empirical parameter selection results in dissipative QRC.

The following simple algebraic criterion characterizes when a QMS is relaxing with a faithful steady state.

\begin{theorem}[Algebraic criterion for relaxation {\citep[Theorem~5.2]{zhang2024criteria}}]\label{thm:part2-algebraic-criterion}
Let $\Phi={\{\Phi_t\}}_{t\ge0}$ be a QMS with jump operators ${\left\{L_\mu\right\}}_\mu$ and Hamiltonian $H$. Define $K=iH+\frac{1}{2}\sum_\mu L_\mu^\ast L_\mu$. Then $\Phi$ is relaxing with a faithful steady state if and only if the associative $\mathbb C$-algebra generated by ${\left\{L_\mu\right\}}_\mu$ and $K$ is the whole operator space, i.e., $\operatorname{alg}_{\mathbb C}({\left\{L_\mu\right\}}_\mu,K)=\Lin(\mathcal X)$.
\end{theorem}

In Ref.~\citep{zhang2024criteria}, the condition on $\Phi$ is stated as irreducibility. By the equivalence result of Ref.~\citep[Theorem~4.2]{fagnola2026irreducibility}, irreducibility is equivalent to primitivity and also to being relaxing with a faithful steady state.

We now apply this criterion to the proposed QRC.

\begin{proposition}[Algebraic verification for the XYZ model]\label{prop:part2-proposed-reservoir-algebra}
Let \(G=(V,E)\) be an undirected simple graph with \(V=\{1,\ldots,N\}\), and let \(\mathcal X={(\mathbb C^2)}^{\otimes N}\). We consider the inhomogeneous anisotropic XYZ Hamiltonian
\begin{equation}
\begin{aligned}
  H
  ={}&
  \sum_{\{j,k\}\in E}
  \Bigl(
    J_{jk}^{x} X_j X_k
    +
    J_{jk}^{y} Y_j Y_k
    +
    J_{jk}^{z} Z_j Z_k
  \Bigr)
  \\
  &+
  \sum_{j=1}^{N}
  \Bigl(
    h_j^{x}X_j
    +
    h_j^{y}Y_j
    +
    h_j^{z}Z_j
  \Bigr),
\end{aligned}
\label{eq:general-anisotropic-heisenberg}
\end{equation}
where all coefficients are real, and we assume that ${\left(h_j^{x}\right)}^2+{\left(h_j^{y}\right)}^2>0$ for every $j\in V$. For every site $j$, let the jump operators be $L_j^{\mathrm{ad}}=\sqrt{\gamma^{\mathrm{ad}}_j}\,\sigma_j^-$ and $L_j^{\mathrm{pd}}=\sqrt{\gamma^{\mathrm{pd}}_j}\,Z_j$, with $\gamma^{\mathrm{ad}}_j\geq0$ and $\gamma^{\mathrm{pd}}_j>0$. Define $K=iH+\frac{1}{2}\sum_\mu L_\mu^\ast L_\mu$, and let $\mathcal A=\operatorname{alg}_{\mathbb C} \left(K,{\{L_j^{\mathrm{ad}},L_j^{\mathrm{pd}}\}}_{j=1}^{N}\right)$. Then $\mathcal A=\Lin(\mathcal X)$.
\end{proposition}

\begin{proof}
See Appendix~\ref{app:proposed-reservoir-algebra}.
\end{proof}

This proposition requires positive pure dephasing rates at every site but allows all amplitude damping rates to vanish. Note that, when the input is encoded solely through the Hamiltonian and pure dephasing is the only decoherence mechanism (i.e., $\gamma^{\mathrm{ad}}_j=0$ for every $j$), the maximally mixed state $I_{2^N}/2^N$ is a faithful stationary state for any Hamiltonian, i.e., $L(I_{2^N}/2^N)=0$. Consequently, whenever the QMS is relaxing in this setting, it converges to the unique faithful steady state \(I_{2^N}/2^N\), independently of the input. This common steady state leads to an asymptotic loss of input dependence even for time-varying inputs, which may limit the reservoir’s ability to process long input sequences~\citep{martinezpena2023finite}. Moreover, Ref.~\citep{yoshida2024uniqueness} considers an XYZ model with local pure dephasing in the absence of the input field in the $xy$ plane, and shows that the dynamics restricted to certain symmetry sectors is irreducible.

Assuming that the local driving fields corresponding to each admissible input are nonzero, it follows from Proposition~\ref{prop:part2-proposed-reservoir-algebra}, Theorems~\ref{thm:part2-algebraic-criterion} and~\ref{thm:part2-qms-equivalence}, and the general results of Subsec.~\ref{subsec:general-finite-dimensional-qrc} that the proposed QRC satisfies the ESP and the FMP. Having established the ESP and the FMP, we next examine how the stationary state depends on the input. The following lemma extends the result in Ref.~\citep[Lemma~1]{sannia2024dissipation} to inhomogeneous anisotropic XYZ Hamiltonians with local amplitude damping and pure dephasing.

\begin{lemma}[Hamiltonian dependence of stationary states]
\label{lem:hamiltonian-stationary-states}
Let $H_1$ and $H_2$ be Hamiltonians of the form in Eq.~\eqref{eq:general-anisotropic-heisenberg} on $\mathcal X=(\mathbb C^2)^{\otimes N}$. Let $J_{jk}^{\alpha}$ denote their common coupling coefficients and $h_{\ell,j}^{\alpha}$ their respective local-field coefficients, where $\ell=1,2$ and $\alpha=x,y,z$. Set $\Delta h_j^\alpha=h_{2,j}^\alpha-h_{1,j}^\alpha$ and $\Delta H=H_2-H_1$. Let $\Phi_1$ and $\Phi_2$ be QMSs with Hamiltonians $H_1$ and $H_2$, respectively, and the same set of jump operators $\{L_j^{\mathrm{ad}},L_j^{\mathrm{pd}}\}_{j=1}^{N}$, where $L_j^{\mathrm{ad}}=\sqrt{\gamma_j^{\mathrm{ad}}}\,\sigma_j^-$, $L_j^{\mathrm{pd}}=\sqrt{\gamma_j^{\mathrm{pd}}}\,Z_j$, $\gamma_j^{\mathrm{ad}}>0$, and $\gamma_j^{\mathrm{pd}}\geq0$ for all $j$. Assume that $\Phi_1$ and $\Phi_2$ admit unique stationary states $\rho_1$ and $\rho_2$, respectively. Then the following statements hold.
\begin{enumerate}
 \item[\textnormal{(i)}]
 $\rho_1=\rho_2$ if and only if $[\Delta H,\rho_1]=0$.

 \item[\textnormal{(ii)}]
 If $\Delta h_j^z=0$ and
 $(\Delta h_j^x)^2+(\Delta h_j^y)^2>0$ for every $j$,
 then $\rho_1\neq\rho_2$.

\end{enumerate}
\end{lemma}
\begin{proof}
See Appendix~\ref{app:proof-hamiltonian-stationary-states}.
\end{proof}

Lemma~\ref{lem:hamiltonian-stationary-states} provides conditions under which different local fields yield different stationary states. In particular, $x$- or $y$-drivings that differ at every site yield distinct stationary states. Hence, under suitable input encoding, the proposed quantum
reservoir has distinct stationary states for distinct inputs. We next turn to the effect of introducing pure dephasing at a fixed input. The following lemma characterizes when the stationary state under amplitude damping alone is preserved.

\begin{lemma}[Effect of pure dephasing on stationary states]
\label{lem:ad-pd-stationary-state}
Let $H$ be a Hamiltonian on $\mathcal X=(\mathbb C^2)^{\otimes N}$, and let $\Phi_1$ and $\Phi_2$ be QMSs with the common Hamiltonian $H$ and respective sets of jump operators $\{L_j^{\mathrm{ad}}\}_{j=1}^{N}$ and $\{L_j^{\mathrm{ad}},L_j^{\mathrm{pd}}\}_{j=1}^{N}$, where $L_j^{\mathrm{ad}}=\sqrt{\gamma_j^{\mathrm{ad}}}\,\sigma_j^-$, $L_j^{\mathrm{pd}}=\sqrt{\gamma_j^{\mathrm{pd}}}\,Z_j$, $\gamma_j^{\mathrm{ad}}>0$, and $\gamma_j^{\mathrm{pd}}\geq 0$ for all $j$. Assume that $\Phi_1$ and $\Phi_2$ admit unique stationary states $\rho_1$ and $\rho_2$, respectively. Set $\rho_0=(\sigma^-\sigma^+)^{\otimes N}$, where $\sigma^-$ and $\sigma^+$ denote the single-qubit lowering and raising operators, respectively. Then the following statements hold.
\begin{enumerate}
 \item[\textnormal{(i)}]
 $\rho_1=\rho_2$ if and only if $[\rho_1,Z_j]=0$ for every $j$ such that $\gamma_j^{\mathrm{pd}}>0$.

 \item[\textnormal{(ii)}]
 If $\gamma_j^{\mathrm{pd}}>0$ for all $j$, then $\rho_1=\rho_2$ if and only if $[H,\rho_0]=0$. In this case, $\rho_1=\rho_2=\rho_0$.

 \item[\textnormal{(iii)}]
 Under the assumption of \textnormal{(ii)}, if $H$ is given by Eq.~\eqref{eq:general-anisotropic-heisenberg}, then $\rho_1=\rho_2$ if and only if $h_j^x=h_j^y=0$ for all $j$ and $J_{jk}^x=J_{jk}^y$ for all $\{j,k\}\in E$.
\end{enumerate}
\end{lemma}

\begin{proof}
See Appendix~\ref{app:proof-ad-pd-stationary-state}.
\end{proof}

Lemma~\ref{lem:ad-pd-stationary-state} shows that introducing pure dephasing at every site alters the stationary state under amplitude damping alone whenever $h_j^x$ or $h_j^y$ is nonzero for some $j$, or $J_{jk}^x\neq J_{jk}^y$ for some $\{j,k\}\in E$. Consequently, pure dephasing can alter the input-dependent steady states and dynamical properties of the proposed quantum reservoir, which can in turn affect its input--output response and computational performance.

\section{Memory and nonlinearity}
\label{sec:memory-nonlinearity}

For general RC systems, memory and nonlinearity are two fundamental computational properties~\citep{tanaka2019recent}. Memory characterizes the extent to which the current reservoir state retains information about past inputs, whereas nonlinearity characterizes the ability of the reservoir dynamics to transform the input history into nonlinear features accessible to a linear readout. A systematic measure of both properties is the information processing capacity, introduced in Ref.~\citep{dambre2012information}. Intuitively, IPC evaluates how accurately linear readouts of the reservoir state can reconstruct different functions of past inputs. By organizing these target functions according to their delays and nonlinearity degrees, IPC reveals how the computational resources of a reservoir are distributed between memory and nonlinear processing.

This section proceeds in two parts. In Subsec.~\ref{subsec:ipc-definition}, we introduce the definition and basic properties of IPC and explain how it is computed for an RC system. In Subsec.~\ref{subsec:ipc-parameters}, we analyze how the effective parameters identified in Subsec.~\ref{subsec:qrc-reservoir} affect the IPC of the proposed QRC. Overall, amplitude damping and pure dephasing shape the IPC in a partially independent, non-additive, and degree-dependent manner, highlighting the benefit of fine-tuning both decoherence rates for individual tasks.

\subsection{Information processing capacity}
\label{subsec:ipc-definition}

Following Ref.~\citep{dambre2012information}, we first formulate IPC in an abstract Hilbert space setting.

\begin{definition}[Information processing capacity]\label{def:ipc}
Let $\mathcal H$ be a real Hilbert space, let $\mathcal R\subseteq\mathcal H$ be a closed linear subspace called the readout space, and let $P_{\mathcal R}$ denote the orthogonal projection onto $\mathcal R$. For any nonzero $z\in\mathcal H$, the capacity of $z$ with respect to $\mathcal R$ is $C_{\mathcal R}(z):=\|P_{\mathcal R}z\|_{\mathcal H}^{2}/ \|z\|_{\mathcal H}^{2}$. A closed linear subspace $\mathcal T\subseteq\mathcal H$ is called a target space. Let $\mathcal Z={\{z_\alpha\}}_{\alpha\in\mathcal I}$ be an orthonormal basis of $\mathcal T$. The family $\mathbf C(\mathcal R;\mathcal Z) :={\left(C_{\mathcal R}(z_\alpha)\right)}_{\alpha\in\mathcal I}$ is called the IPC profile with respect to $\mathcal Z$. The total IPC associated with $\mathcal T$ is $C_{\mathrm{tot}}(\mathcal R;\mathcal T) :=\sum_{\alpha\in\mathcal I}C_{\mathcal R}(z_\alpha)$. When $\mathcal R$, $\mathcal T$, and $\mathcal Z$ are clear from context, we write $C_\alpha:=C_{\mathcal R}(z_\alpha)$, $\mathbf C:={\left(C_\alpha\right)}_{\alpha\in\mathcal I}$, and $C_{\mathrm{tot}}:=\sum_{\alpha\in\mathcal I}C_\alpha$.
\end{definition}

\begin{proposition}[Basic properties of IPC]\label{prop:ipc-properties}
Assume that the readout space $\mathcal R$ has finite dimension $r$. Then the following statements hold.
\begin{enumerate}
  \item[\textnormal{(i)}] For every nonzero $z\in\mathcal H$, $0\le C_{\mathcal R}(z)\le1$. Moreover, $C_{\mathcal R}(z)=1$ if and only if $z\in\mathcal R$, and $C_{\mathcal R}(z)=0$ if and only if $z\in\mathcal R^\perp$, where ${\left(\cdot\right)}^\perp$ denotes the orthogonal complement.

  \item[\textnormal{(ii)}] The total IPC is independent of the orthonormal basis $\mathcal Z$ of $\mathcal T$. More precisely, for any orthonormal basis ${\{e_j\}}_{j=1}^{r}$ of $\mathcal R$, $C_{\mathrm{tot}}(\mathcal R;\mathcal T) =\sum_{j=1}^{r}\|P_{\mathcal T}e_j\|_{\mathcal H}^{2}$.

  \item[\textnormal{(iii)}] We have $C_{\mathrm{tot}}(\mathcal R;\mathcal T)\le r$, with equality if and only if $\mathcal R\subseteq\mathcal T$. If $\mathcal T$ is finite-dimensional, then $C_{\mathrm{tot}}(\mathcal R;\mathcal T) \le\min\{r,\dim\mathcal T\}$. In particular, $C_{\mathrm{tot}}(\mathcal R;\mathcal H)=r$.

  \item[\textnormal{(iv)}] If $\mathcal R_1\subseteq\mathcal R_2$, then $C_{\mathcal R_1}(z)\le C_{\mathcal R_2}(z)$ for every nonzero $z\in\mathcal H$, and $C_{\mathrm{tot}}(\mathcal R_1;\mathcal T) \le C_{\mathrm{tot}}(\mathcal R_2;\mathcal T)$.
\end{enumerate}
\end{proposition}

\begin{proof}
See Appendix~\ref{app:ipc-properties}.
\end{proof}

We next specialize Definition~\ref{def:ipc} to a general RC system. Let $D_{N_{\mathrm{u}}}\subset\mathbb R^{N_{\mathrm{u}}}$ and $D_{N_{\mathrm{x}}}\subset\mathbb R^{N_{\mathrm{x}}}$ be compact input and state sets, respectively. Consider an RC system with the ESP, consisting of a reservoir map $F:D_{N_{\mathrm{x}}}\times D_{N_{\mathrm{u}}} \to D_{N_{\mathrm{x}}}$ and a linear readout map $h:D_{N_{\mathrm{x}}}\to\mathbb R^{N_{\mathrm{y}}}$. The ESP induces a unique reservoir functional $U^F:D_{N_{\mathrm{u}}}^{\mathbb Z_-}\to D_{N_{\mathrm{x}}}$, where $\mathbb Z_-=\{0,-1,-2,\ldots\}$ and $U^F(\ldots,\mathbf u_{-1},\mathbf u_0)$ is the current reservoir state generated by the left-infinite input history $(\ldots,\mathbf u_{-1},\mathbf u_0)$.

Let $\mu$ be a Borel probability measure on $D_{N_{\mathrm{u}}}$, and suppose that the inputs are independent and identically distributed according to $\mu$. We equip the input-history space $D_{N_{\mathrm{u}}}^{\mathbb Z_-}$ with the product $\sigma$-algebra $\mathcal F={\mathcal B(D_{N_{\mathrm{u}}})}^{\otimes\mathbb Z_-}$ and the product probability measure $\mathbb P_\mu=\mu^{\otimes\mathbb Z_-}$. We take the ambient Hilbert space in Definition~\ref{def:ipc} to be the real $L^2$ space $\mathcal H=L^2(D_{N_{\mathrm{u}}}^{\mathbb Z_-}, \mathcal F,\mathbb P_\mu;\mathbb R)$. For each $i\in\{1,\ldots,N_{\mathrm{x}}\}$, let $\pi_i:D_{N_{\mathrm{x}}}\to\mathbb R$ denote the $i$th coordinate projection and define $U_i^F=\pi_i\circ U^F$. Since $D_{N_{\mathrm{x}}}$ is compact, each $U_i^F$ is bounded and hence belongs to $\mathcal H$.

For a scalar-output linear readout $h:D_{N_{\mathrm{x}}}\to\mathbb R$, there exists $\mathbf w\in\mathbb R^{N_{\mathrm{x}}}$ such that $h(\mathbf x)=\mathbf w^\top\mathbf x$. Its associated output functional is therefore $U^F_h=h\circ U^F=\sum_{i=1}^{N_{\mathrm{x}}} w_i U_i^F$. Thus, the scalar functions realizable by linear readouts form the readout space $\mathcal R:=\operatorname{span}\{U_1^F,\ldots,U_{N_{\mathrm{x}}}^F\} \subseteq\mathcal H$. We choose the target space $\mathcal T=\mathcal H$ and let $\mathcal Z={\{z_\alpha\}}_{\alpha\in\mathcal I}$ be an orthonormal basis of $\mathcal T$. The capacity of $z_\alpha$ is $C_\alpha=\|P_{\mathcal R}z_\alpha\|_{\mathcal H}^{2} =1-\inf_{\mathbf w\in\mathbb R^{N_{\mathrm{x}}}} \|z_\alpha-\mathbf w^\top U^F\|_{\mathcal H}^{2}$. The IPC profile is ${\left(C_\alpha\right)}_{\alpha\in\mathcal I}$, and its total is $C_{\mathrm{tot}}=\sum_{\alpha\in\mathcal I}C_\alpha =\dim\mathcal R\le N_{\mathrm{x}}$ by Proposition~\ref{prop:ipc-properties}.

A standard choice of orthonormal basis for $\mathcal T$ is the Legendre basis~\citep{dambre2012information}. Ref.~\citep{kubota2021unifying} summarizes various input distributions and their corresponding orthonormal polynomial bases. Suppose that $D_{N_{\mathrm{u}}}={[-1,1]}^{N_{\mathrm{u}}}$ and that $\mu=\lambda^{\otimes N_{\mathrm{u}}}$, where $\mathrm{d}\lambda(x)=\frac{1}{2}\mathrm{d}x$ is the uniform probability measure on $[-1,1]$. Let $p_k$ be the Legendre polynomial of degree $k$ and set $\ell_k(x)=\sqrt{2k+1}\,p_k(x)$. Then ${\{\ell_k\}}_{k\in\mathbb N_0}$ is an orthonormal basis of $L^2([-1,1],\lambda)$.

For an input history $\mathbf u=(\ldots,\mathbf u_{-1},\mathbf u_0) \in D_{N_{\mathrm{u}}}^{\mathbb Z_-}$, let $u_{-\tau,j}$ denote the $j$th component of $\mathbf u_{-\tau}$. Let $\mathfrak A$ be the set of all finitely supported multi-indices $\boldsymbol\alpha ={\left(\alpha_{\tau,j}\right)}_{ \tau\in\mathbb N_0,\,1\le j\le N_{\mathrm{u}}}$ with $\alpha_{\tau,j}\in\mathbb N_0$. For each $\boldsymbol\alpha\in\mathfrak A$, define
\begin{equation}
  z_{\boldsymbol\alpha}(\mathbf u)
  :=\prod_{\tau=0}^{\infty}\prod_{j=1}^{N_{\mathrm{u}}}
  \ell_{\alpha_{\tau,j}}(u_{-\tau,j}).
  \label{eq:legendre-history-basis}
\end{equation}
Since $\ell_0=1$ and $\boldsymbol\alpha$ has finite support, the product in Eq.~\eqref{eq:legendre-history-basis} contains only finitely many nonconstant factors. Independence of the input coordinates and time steps implies that ${\{z_{\boldsymbol\alpha}\}}_{\boldsymbol\alpha\in\mathfrak A}$ is an orthonormal basis of $\mathcal T$.

The total polynomial degree of $z_{\boldsymbol\alpha}$ is $|\boldsymbol\alpha| =\sum_{\tau=0}^{\infty}\sum_{j=1}^{N_{\mathrm{u}}} \alpha_{\tau,j}$. Accordingly, the degree-$p$ IPC is $C^{(p)}=\sum_{|\boldsymbol\alpha|=p}C_{\boldsymbol\alpha}$. The degree-zero term corresponds to the constant target, the degree-one IPC quantifies linear memory, and the degree-$p$ IPCs with $p\ge2$ quantify nonlinear information-processing capacity.

In practical application, only a finite input sequence is available. After discarding an initial washout period, suppose that we retain $M$ time steps with reservoir states $\mathbf x[1],\ldots,\mathbf x[M]\in D_{N_{\mathrm{x}}}$. The  Hilbert space $\mathcal H$ is then replaced by the empirical Hilbert space $\mathcal H_M=\mathbb R^M$ with inner product $\langle\mathbf a,\mathbf b\rangle_M=M^{-1}\mathbf a^\top\mathbf b$ and norm $\|\mathbf a\|_M=\sqrt{\langle\mathbf a,\mathbf a\rangle_M}$. The state data matrix and sampled target vector are given by $X_M=[\mathbf x[1]\ \mathbf x[2]\ \cdots\ \mathbf x[M]] \in\mathbb R^{N_{\mathrm{x}}\times M}$ and $\mathbf z={\left(z[1],\ldots,z[M]\right)}\in\mathbb R^M$, respectively.

A linear readout with weight vector $\mathbf w\in\mathbb R^{N_{\mathrm{x}}}$ produces the sampled output $X_M^\top\mathbf w\in\mathcal H_M$. Hence, the empirical readout space is $\mathcal R_M=\operatorname{row}(X_M)\subseteq\mathcal H_M$. The orthogonal projector onto $\mathcal R_M$ is $P_{\mathcal R_M}=X_M^\dagger X_M$, where ${\left(\cdot\right)}^\dagger$ denotes the Moore--Penrose pseudoinverse. Consequently, the empirical capacity of a nonzero sampled target $\mathbf z$ is
\begin{equation}
  \begin{aligned}
    \widehat C(\mathbf z)
    &=\frac{\|P_{\mathcal R_M}\mathbf z\|_M^2}
    {\|\mathbf z\|_M^2}\\
    &=1-\frac{\min_{\mathbf w\in\mathbb R^{N_{\mathrm{x}}}}
    \|\mathbf z- X_M^\top\mathbf w\|_M^2}
    {\|\mathbf z\|_M^2}.
  \end{aligned}
  \label{eq:empirical-ipc-capacity}
\end{equation}
The solution of the least squares problem in Eq.~\eqref{eq:empirical-ipc-capacity} is $\mathbf w^\star(\mathbf z)={(X_M^\top)}^\dagger\mathbf z$. For each basis target $z_\alpha$, denote its sampled vector by $\mathbf z_\alpha$ and its empirical capacity by $\widehat C_\alpha=\widehat C(\mathbf z_\alpha)$. Then, the empirical IPC profile is ${\left(\widehat C_\alpha\right)}_{\alpha\in\mathcal I}$, and the empirical total IPC over a finite collection of targets is the sum of their capacities.

\begin{figure*}[!tp]
  \centering
  \widefiguregraphics{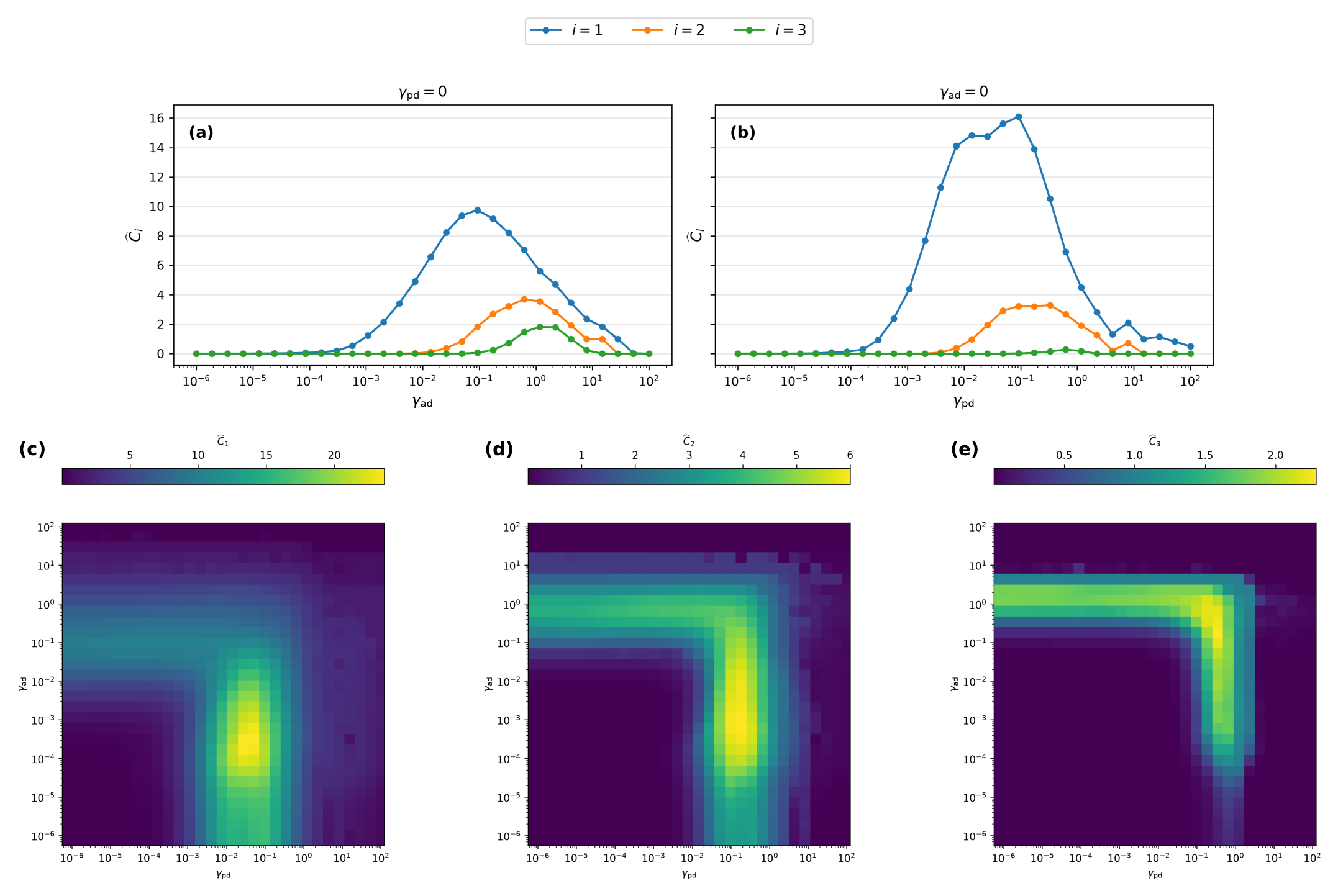}
  \caption{Effects of amplitude damping and pure dephasing on $\widehat C_i$ ($i=1,2,3$) for the 4-qubit QRC. Panels~(a) and (b) show the individual dependence on $\gamma_{\mathrm{ad}}$ and $\gamma_{\mathrm{pd}}$, with $\gamma_{\mathrm{pd}}=0$ and $\gamma_{\mathrm{ad}}=0$, respectively. Panels~(c)--(e) show the joint dependence of $\widehat C_1$, $\widehat C_2$, and $\widehat C_3$, respectively, on $\gamma_{\mathrm{ad}}$ and $\gamma_{\mathrm{pd}}$. In all panels, $\alpha_{\mathrm{in}}=\mathrm{0.01}$, $J=\mathrm{1.0}$, and $h_{\mathrm{z}}=\mathrm{0.2}$.}
  \label{fig:nmc4-gammas}
\end{figure*}

\subsection{Effects of hyperparameters on IPC}
\label{subsec:ipc-parameters}

Direct analysis of the full IPC profile is difficult because it contains infinitely many targets, including interactions among arbitrary delays. Following Ref.~\citep{gotting2025connection}, we quantify the memory performance by the memory capacity at each polynomial degree. Specifically, we consider scalar inputs (i.e., $N_{\mathrm{u}}=1$), denote $u_{-\tau,1}$ by $u_{-\tau}$, and use the Legendre basis introduced above. We then identify $\mathfrak A$ with the set of finitely supported sequences ${\left(\alpha_\tau\right)}_{\tau\in\mathbb N_0}$. For $i\in\mathbb N$ and $\tau\in\mathbb N_0$, define $\boldsymbol\alpha^{(i,\tau)}\in\mathfrak A$ by
\begin{equation*}
  \alpha^{(i,\tau)}_{\tau'}
  :=
  \begin{cases}
    i, & \tau'=\tau,\\
    0, & \text{otherwise}.
  \end{cases}
\end{equation*}
The corresponding target and its capacities are $z_{i,\tau}=z_{\boldsymbol\alpha^{(i,\tau)}} =\ell_i(u_{-\tau})$, $C_{i,\tau}=C_{\boldsymbol\alpha^{(i,\tau)}}$, and $\widehat C_{i,\tau}=\widehat C(\mathbf z_{i,\tau})$, where $\mathbf z_{i,\tau}$ is the sampled vector of $z_{i,\tau}$. For a prescribed maximum delay $\tau_{\max}$, we define the theoretical and empirical degree-$i$ memory capacities as $C_i=\sum_{\tau=1}^{\infty}C_{i,\tau}$ and $\widehat C_i=\sum_{\tau=1}^{\tau_{\max}}\widehat C_{i,\tau}$, respectively. The quantities $C_1$ and $\widehat C_1$ quantify linear memory, whereas $C_i$ and $\widehat C_i$ for $i\ge2$ quantify degree-$i$ nonlinear memory. Since the sum defining $C_i$ includes only targets that depend on a single delayed input, $C_i\le C^{(i)}$.

For the proposed QRC, we investigate the effects of the five effective hyperparameters on memory performance using $\widehat C_1$, $\widehat C_2$, and $\widehat C_3$. IPC analyses reveal a trade-off between memory and nonlinearity in generic RC systems: increasing nonlinear processing capacity generally comes at the expense of linear memory, and vice versa~\citep{dambre2012information}. Since $\widehat C_1$ quantifies linear memory, whereas $\widehat C_2$ and $\widehat C_3$ quantify second- and third-degree nonlinear memory, respectively, their joint analysis also characterizes how each hyperparameter affects the nonlinear processing capability of the systems.

To ensure that the empirical capacities $\widehat C_i$ accurately approximate $C_i$, we choose sufficiently large values for both the maximum delay $\tau_{\max}$ and the input sequence length. We verify numerical convergence by confirming that the resulting $\widehat C_i$ values change negligibly when either quantity is further increased. Note that, for the reservoir subject to pure dephasing alone, the associated filter is constant and hence $C_i=0$. Only in this case do we focus on the empirical capacities $\widehat C_i$ computed from finite input sequences without requiring convergence to $C_i$. Further details of the convergence analysis are provided in Appendix~\ref{app:ipc-convergence}. Specifically, for the 4-qubit QRC, we set $\tau_{\max}=30$ and set the input sequence length to $52500$. We discard the first $2500$ time steps as washout and split the remaining $50000$ time steps into $25000$ training and $25000$ test time steps. The empirical capacities are then computed out of sample on the test segment.

The effects of amplitude damping and pure dephasing on $\widehat C_i$ ($i=1,2,3$) for the 4-qubit QRC are shown in Fig.~\ref{fig:nmc4-gammas}. Panels~(a) and (b) show the individual dependence of $\widehat C_i$ on $\gamma_{\mathrm{ad}}$ and $\gamma_{\mathrm{pd}}$, with $\gamma_{\mathrm{pd}}=0$ and $\gamma_{\mathrm{ad}}=0$, respectively. In each panel, the three $\widehat C_i$ quantities exhibit their respective active regions over the corresponding decoherence rate. Within these regions, the three quantities can be collectively characterized by a broadly hump-shaped pattern, generally increasing at lower decoherence rates and decreasing at higher rates. Outside their respective active regions, the quantities are strongly suppressed. The linear and nonlinear memory capacities of the reservoir can therefore be tuned by varying the corresponding decoherence rate within these active regions. The two decoherence mechanisms nevertheless produce distinct responses: pure dephasing produces a stronger response in $\widehat C_1$ than amplitude damping, whereas its response in $\widehat C_3$ is comparatively weaker. Moreover, the active regions differ in their widths and locations under the two decoherence mechanisms.

The joint dependence of $\widehat C_i$ on $\gamma_{\mathrm{ad}}$ and $\gamma_{\mathrm{pd}}$ is shown in panels~(c)--(e). In each panel, the active region of $\widehat C_i$ exhibits an approximately L-shaped pattern. Its two arms correspond to regions in which one decoherence rate is dominant relative to the other, whereas the two rates are comparable near the corner. As either decoherence rate decreases, the dependence of $\widehat C_i$ on the other rate smoothly approaches the corresponding single-decoherence dependence shown in panels~(a) and (b). This behavior indicates that amplitude damping and pure dephasing can tune the linear and nonlinear memory capacities with partial independence. However, the peak region of the $\widehat C_i$ distribution does not necessarily coincide with the corner of the L-shaped active region, suggesting a non-additive response to the two decoherence mechanisms. Moreover, $\widehat C_1$, $\widehat C_2$, and $\widehat C_3$ differ in both magnitude and distribution within their active regions. Consequently, decoherence rates that enhance one quantity do not necessarily enhance the others to the same extent, indicating a degree-dependent response to decoherence.

Together, the partial independence, non-additivity, and degree dependence of the responses highlight the broader value of controlling both amplitude damping and pure dephasing. The partial independence of the two mechanisms provides a degree of operational flexibility and broadens the range of viable operating conditions. Moreover, owing to the non-additivity and degree dependence of the responses, joint control of the two decoherence mechanisms allows the relative magnitudes of $\widehat C_1$, $\widehat C_2$, and $\widehat C_3$ to be regulated, thereby providing access to $\widehat C_i$ profiles that may not be obtained by tuning either decoherence rate alone.

\begin{figure*}[!tp]
  \centering
  \widefiguregraphics{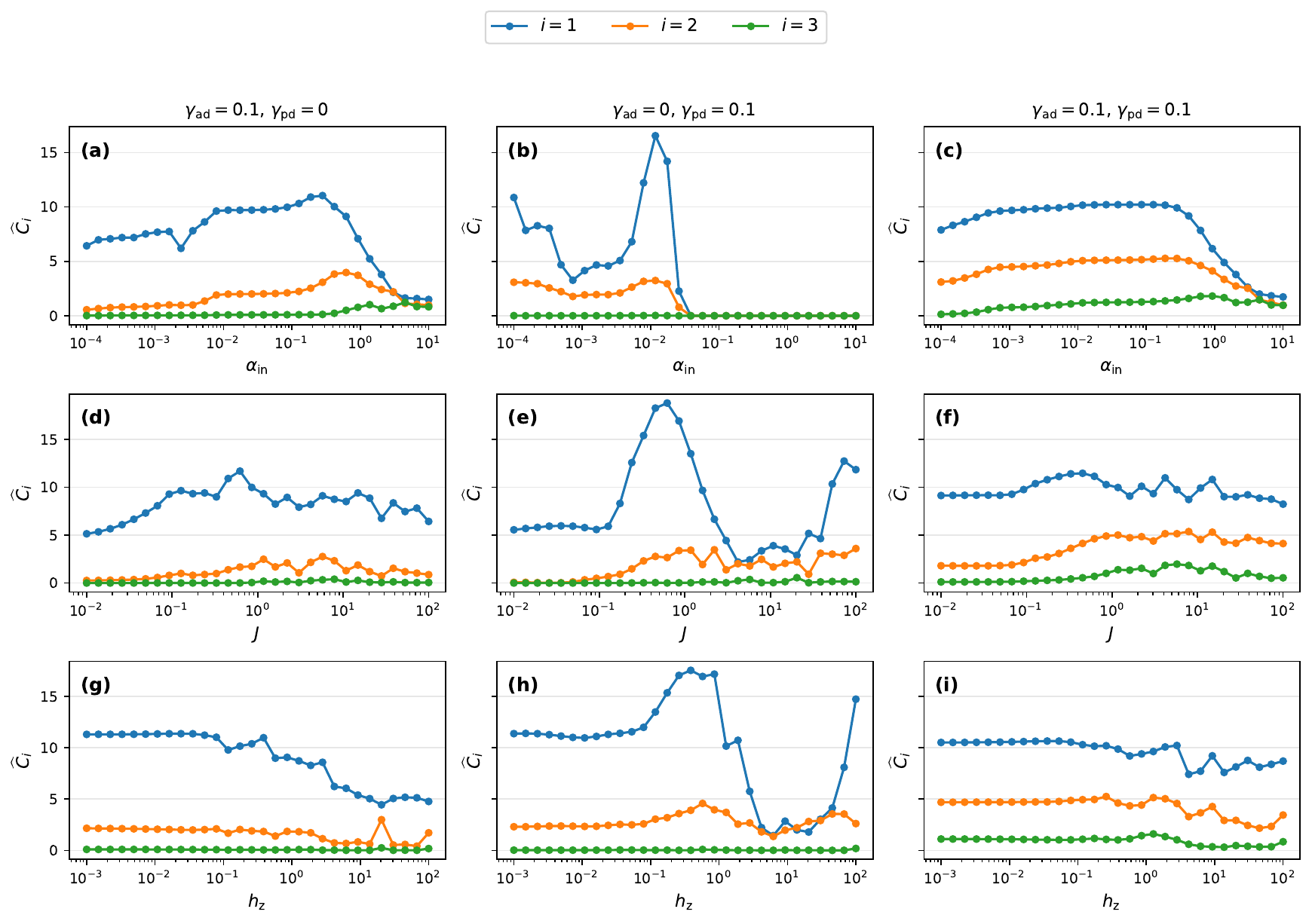}
  \caption{Effects of input scaling $\alpha_{\mathrm{in}}$, coupling strength $J$, and transverse field strength $h_{\mathrm{z}}$ on $\widehat C_i$ ($i=1,2,3$) for the 4-qubit QRC. Panels (a)--(c), (d)--(f), and (g)--(i) show the dependence of $\widehat C_i$ on $\alpha_{\mathrm{in}}$, $J$, and $h_{\mathrm{z}}$, respectively. The first, second, and third columns correspond to amplitude damping only, $(\gamma_{\mathrm{ad}},\gamma_{\mathrm{pd}})=(0.1,0)$; pure dephasing only, $(0,0.1)$; and simultaneous amplitude damping and pure dephasing, $(0.1,0.1)$, respectively. In each panel, the two hyperparameters not being varied are fixed at their corresponding values in the reference setting, $(\alpha_{\mathrm{in}},J,h_{\mathrm{z}})=(\mathrm{0.01},\mathrm{1.0},\mathrm{0.2})$.}
  \label{fig:nmc4-reservoir}
\end{figure*}

We also study the dependence of $\widehat C_i$ on other QRC hyperparameters, namely input scaling $\alpha_{\mathrm{in}}$, coupling strength $J$, and transverse field strength $h_{\mathrm{z}}$, as shown in Fig.~\ref{fig:nmc4-reservoir}. The three rows show the dependence on $\alpha_{\mathrm{in}}$, $J$, and $h_{\mathrm{z}}$ from top to bottom, while the three columns correspond to $(\gamma_{\mathrm{ad}},\gamma_{\mathrm{pd}})=(0.1,0)$, $(0,0.1)$, and $(0.1,0.1)$ from left to right. Under amplitude damping alone, all three hyperparameters can affect the $\widehat C_i$ values in different patterns. Specifically, in panel~(a), the capacities increase gradually with $\alpha_{\mathrm{in}}$ before decreasing rapidly in the strong-input range. Meanwhile, the capacities show limited fluctuations with $J$ in panel~(d), whereas they generally decrease with increasing $h_{\mathrm{z}}$ in panel~(g). Thus, the effects of the three hyperparameters remain moderate across broad hyperparameter ranges, except for the strong-input range.

By contrast, under pure dephasing alone, the capacities exhibit intense variations with all three hyperparameters. In panel~(b), the capacities vary substantially in the weak-input range and are rapidly suppressed in the cutoff region. In panels~(e) and (h), the capacities remain stable at small $J$ and $h_{\mathrm{z}}$ but exhibit strong variations as these hyperparameters increase. The capacities are therefore more sensitive to all three hyperparameters under pure dephasing alone.

When both decoherence mechanisms are present, the capacities remain stable over broad ranges of the three hyperparameters compared with their behavior under the single-decoherence settings. In addition, as in panel~(a), the capacities decrease rapidly in the strong-input range, as shown in panel~(c). These results indicate that the coexistence of amplitude damping and pure dephasing generally renders the linear and nonlinear memory capacities largely immune to the three hyperparameters. This broad robustness provides a foundation for the subsequent analysis mainly focusing on the decoherence rates.

\section{Benchmark tasks}
\label{sec:benchmark-tasks}

To assess the proposed QRC, we first examine its performance on standard synthetic benchmark tasks. Ref.~\citep{wringe2025reservoir} summarizes benchmark tasks commonly used to evaluate RC systems. In Subsec.~\ref{subsec:synthetic-benchmark-tasks}, we introduce the four families of synthetic benchmark tasks considered in this work, covering computation, imitation, and prediction. In Subsec.~\ref{subsec:synthetic-benchmark-results}, we present the corresponding experimental results. The results show that the proposed QRC achieves promising performance across all the considered benchmarks.

\subsection{Synthetic benchmark tasks}
\label{subsec:synthetic-benchmark-tasks}

We evaluate the proposed QRC on four families of synthetic benchmark tasks: temporal exclusive OR (XOR) computation, nonlinear autoregressive moving average (NARMA) model imitation, Mackey--Glass system (MGS) prediction, and Lorenz system prediction. Throughout this section, $L\in\mathbb N$ denotes the input and target sequence length.

\paragraph*{Temporal XOR.} Let $\tau\in\{1,\ldots,L-1\}$ be the temporal delay. The input entries are independently sampled according to $u[n]\overset{\mathrm{i.i.d.}}{\sim} \operatorname{Bernoulli}\!\left(1/2\right)$ for $n=1,\ldots,L$. Let $\oplus$ denote the XOR operation, which for $a,b\in\{0,1\}$ satisfies $a\oplus b=(a+b)\bmod 2=a+b-2ab$. The target sequence is then defined by the delayed XOR rule as
\begin{equation*}
  d[n]
  =
  \begin{cases}
    0, & 1\le n\le\tau,\\
    u[n]\oplus u[n-\tau], & \tau+1\le n\le L.
  \end{cases}
\end{equation*}
\paragraph*{NARMA.} Let $q\in\{1,\ldots,L-1\}$ be the memory order and let $\beta>0$ be the input scale. The input entries are independently sampled according to $u[n]\overset{\mathrm{i.i.d.}}{\sim} \operatorname{Uniform}(0,\beta)$ for $n=1,\ldots,L$. The target is initialized by $d[n]=0$ for $1\le n\le q$. For $q+1\le n\le L$, it is generated recursively by the $q$th-order NARMA model,
\begin{equation*}
  \begin{aligned}
    d[n]
    &=a_1d[n-1]
    +a_2d[n-1]\sum_{j=n-q}^{n-1}d[j]\\
    &\quad+a_3u[n-q+1]u[n]+a_4,
  \end{aligned}
\end{equation*}
where $a_1,a_2,a_3,a_4$ are real-valued model parameters.

\paragraph*{Mackey--Glass system.} The Mackey--Glass system is governed by the scalar nonlinear delay differential equation
\begin{equation*}
  \frac{\mathrm{d}x(t)}{\mathrm{d}t}
  =a\frac{x(t-\tau)}{1+{\bigl[x(t-\tau)\bigr]}^{p}}
  -bx(t),
\end{equation*}
where $a,b,p,$ and $\tau$ are real-valued parameters. The numerical solution of this equation gives the input sequence ${\left\{u[n]\right\}}_{n=1}^{L}$. For a prediction horizon $K\in\mathbb N$, the target is the $K$-step-ahead value of the same solution, $d[n]=u[n+K]$.

\paragraph*{Lorenz system.} The Lorenz system is governed by the three-dimensional nonlinear differential equations
\begin{equation*}
  \begin{aligned}
    \frac{\mathrm{d}x(t)}{\mathrm{d}t}
    &=\sigma\bigl(y(t)-x(t)\bigr),\\
    \frac{\mathrm{d}y(t)}{\mathrm{d}t}
    &=x(t)\bigl(r-z(t)\bigr)-y(t),\\
    \frac{\mathrm{d}z(t)}{\mathrm{d}t}
    &=x(t)y(t)-b z(t),
  \end{aligned}
\end{equation*}
where $\sigma,r,$ and $b$ are real-valued parameters. The numerical solution of this system gives the three-dimensional input sequence ${\left\{\mathbf u[n]\right\}}_{n=1}^{L}$. For a prediction horizon $K\in\mathbb N$, the target is the $K$-step-ahead state of the same solution, $\mathbf d[n]=\mathbf u[n+K]$.

The parameter settings for each task are summarized in Table~\ref{tab:synthetic-task-parameters}.

\begin{table}[!tbp]
  \centering
  \caption{Parameter settings of the synthetic benchmark tasks.}%
  \label{tab:synthetic-task-parameters}
  \tablelayout[4pt]{1.2}
  \fittable{%
    \begin{tabular}{@{}ll@{}}
      \toprule
      Task & Parameters \\
      \midrule
      Temporal XOR & $\tau=1$ \\
      NARMA-5 & \makecell[tl]{$q=5$ \\ $\bigl(a_1,a_2,a_3,a_4\bigr)=\bigl(0.3,0.05,1.5,0.1\bigr)$ \\ $\beta=0.5$} \\
      NARMA-10 & \makecell[tl]{$q=10$ \\ $\bigl(a_1,a_2,a_3,a_4\bigr)=\bigl(0.3,0.05,1.5,0.1\bigr)$ \\ $\beta=0.5$} \\
      MGS-17 & \makecell[tl]{$\bigl(a,b,p,\tau\bigr)=\bigl(0.2,0.1,10,17\bigr)$ \\ $K=84$} \\
      Lorenz & \makecell[tl]{$\bigl(\sigma,r,b\bigr)=\bigl(10,28,8/3\bigr)$ \\ $K=1$} \\
      \bottomrule
    \end{tabular}%
  }
\end{table}

\subsection{Results}
\label{subsec:synthetic-benchmark-results}

For every task, we first apply min-max rescaling to generated input sequences, mapping their values to the interval $[0,1]$. The rescaled sequences are then divided into washout, training, validation, and test segments of lengths $200$, $1200$, $400$, and $400$, respectively. The ridge regularization strength is set to $\lambda=10^{-6}$ for all tasks. Each experiment is repeated over $20$ independent trials. We evaluate the performance using the normalized root mean squared error (NRMSE). For an output sequence $\mathbf y$ and the corresponding target sequence $\mathbf d$, both of length $M$, the NRMSE is defined as
\begin{equation}
  \operatorname{NRMSE}(\mathbf y,\mathbf d)
  ={\left(
    \frac{\sum_{n=1}^{M}\|\mathbf y[n]-\mathbf d[n]\|^2}
    {\sum_{n=1}^{M}\|\mathbf d[n]-\overline{\mathbf d}\|^2}
  \right)}^{1/2},
  \label{eq:benchmark-nrmse}
\end{equation}
where $\overline{\mathbf d}=M^{-1}\sum_{n=1}^{M}\mathbf d[n]$. Smaller NRMSE values indicate better performance, with zero corresponding to perfect performance. The QRC hyperparameters are selected by minimizing the NRMSE on the validation segment, and the resulting setting is evaluated using the NRMSE on the test segment. 

\begin{table}[!t]
  \centering
  \caption{NRMSE results (mean $\pm$ standard deviation) of the 4- and 6-qubit QRCs across the five benchmark tasks.}
  \label{tab:results-benchmarking}
  \tablelayout{1.08}
  \fittable{%
    \begin{tabular}{@{}llc@{}}
      \toprule
      Dataset & Model & NRMSE \\
      \midrule
      Temporal XOR & \ourscell{QRC (4 qubits)} & \ourscell{\resultcell{4.35E-04}{1.35E-04}} \\
      & \ourscell{QRC (6 qubits)} & \ourscell{\resultcell{2.07E-04}{6.2E-05}} \\
      \midrule
      NARMA-5 & \ourscell{QRC (4 qubits)} & \ourscell{\resultcell{6.59E-02}{1.87E-02}} \\
      & \ourscell{QRC (6 qubits)} & \ourscell{\resultcell{2.25E-02}{4.75E-03}} \\
      \midrule
      NARMA-10 & \ourscell{QRC (4 qubits)} & \ourscell{\resultcell{2.46E-01}{4.39E-02}} \\
      & \ourscell{QRC (6 qubits)} & \ourscell{\resultcell{1.37E-01}{2.80E-02}} \\
      \midrule
      MGS-17 & \ourscell{QRC (4 qubits)} & \ourscell{\resultcell{1.76E-03}{4.80E-04}} \\
      & \ourscell{QRC (6 qubits)} & \ourscell{\resultcell{4.08E-04}{1.18E-04}} \\
      \midrule
      Lorenz & \ourscell{QRC (4 qubits)} & \ourscell{\resultcell{1.74E-02}{4.40E-02}} \\
      & \ourscell{QRC (6 qubits)} & \ourscell{\resultcell{1.69E-04}{1.03E-04}} \\
      \bottomrule
    \end{tabular}%
  }
\end{table}

\begin{table}[!t]
  \centering
  \caption{Optimized hyperparameter settings for the 4- and 6-qubit QRCs across the five benchmark tasks. Each entry is formatted as $\alpha_{\mathrm{in}}/J/h_{\mathrm{z}}/\gamma_{\mathrm{ad}}/\gamma_{\mathrm{pd}}$.}
  \label{tab:param-benchmarking}
  \tablelayout{1.08}
  \fittable{%
    \begin{tabular}{lcc}
      \toprule
      Dataset & QRC (4 qubits) & QRC (6 qubits)\\
      \midrule
      Temporal XOR & \paramcell{2.6}{1.9}{0.29}{1.8}{0.0020} & \paramcell{3.0}{2.2}{0.73}{4.5}{0.0040} \\
      \addlinespace[2pt]
      NARMA-5 & \paramcell{0.19}{0.61}{0.62}{0.030}{0.11} & \paramcell{0.20}{0.82}{0.030}{0.026}{0.076} \\
      \addlinespace[2pt]
      NARMA-10 & \paramcell{0.14}{0.41}{0.60}{0.012}{0.070} & \paramcell{0.11}{0.80}{1.0}{0.0057}{0.062} \\
      \addlinespace[2pt]
      MGS-17 & \paramcell{0.35}{0.13}{0.0}{0.053}{0.0} & \paramcell{0.59}{0.32}{0.0}{0.0056}{0.0} \\
      Lorenz & \paramcell{0.27}{0.31}{0.0}{0.23}{0.0} & \paramcell{0.57}{1.6}{0.020}{0.58}{0.0} \\
      \bottomrule
    \end{tabular}%
  }
\end{table}

\begin{figure*}[!tp]
  \centering
  \widefiguregraphics{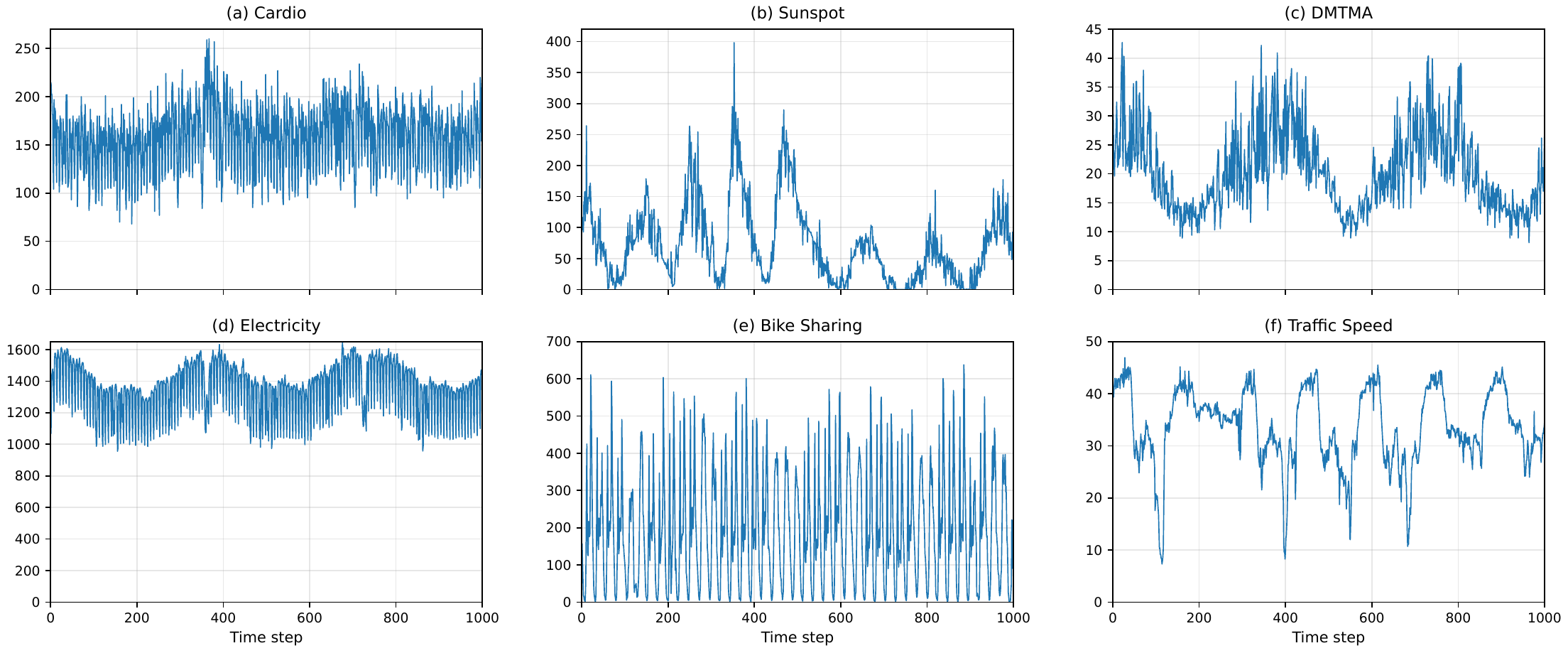}
  \caption{Time-series plots of the six real-world datasets, showing the first 1000 time steps of each dataset.}
  \label{fig:prediction-datasets}
\end{figure*}

The NRMSE results (mean $\pm$ standard deviation) of the 4- and 6-qubit QRCs across the five benchmark tasks are summarized in Table~\ref{tab:results-benchmarking}. For the temporal XOR task, the 4-qubit QRC achieves a remarkably low mean NRMSE of $4.35\times10^{-4}$, demonstrating that the proposed QRC can reliably capture the temporal dependencies underlying the XOR operation using only a few qubits. Increasing the reservoir size to six qubits further reduces the mean NRMSE to $2.07\times10^{-4}$. For the NARMA tasks, namely NARMA-5 and NARMA-10, which demand greater memory and nonlinear processing capabilities at higher orders, the 4-qubit QRC maintains reasonable performance, while the 6-qubit QRC consistently achieves superior performance. For both reservoir sizes, the imitation performance degrades as the NARMA order increases, reflecting the greater difficulty of higher-order tasks. For the MGS-17 and Lorenz prediction tasks, the 4-qubit QRC also demonstrates reasonable performance, whereas increasing the reservoir size to six qubits yields substantially improved performance for both tasks. Overall, these results demonstrate that the proposed QRC achieves promising performance across diverse benchmark tasks and consistently benefits from an increased reservoir size.

The optimized hyperparameter settings for the 4- and 6-qubit QRC configurations across the five benchmark tasks are summarized in Table~\ref{tab:param-benchmarking}. The analysis of the settings is based on Sec.~\ref{subsec:ipc-parameters}.

The optimized hyperparameter settings can be divided into two types, namely single- and dual-decoherence. For MGS-17 and Lorenz, the optimized settings for both the 4- and 6-qubit configurations involve amplitude damping alone, which indicates that effective MGS-17 and Lorenz predictions can be achieved without fine-tuning the memory capacities through joint control of both decoherence mechanisms.

By contrast, the optimized settings for the other benchmark tasks involve both decoherence mechanisms. For temporal XOR, both configurations involve strong input scaling and amplitude-damping-dominated settings. As shown in panel~(c) of Fig.~\ref{fig:nmc4-reservoir}, strong input scaling is associated with reduced memory capacities, which is consistent with the temporal delay setting and nonlinear computational structure of temporal XOR.

For NARMA-5 and NARMA-10, both configurations involve pure-dephasing-dominated settings. As shown in panels~(c)--(e) of Fig.~\ref{fig:nmc4-gammas}, increasing the NARMA order from 5 to 10 shifts the capacities associated with the optimized setting toward a larger $\widehat C_1$, a comparable $\widehat C_2$, and a smaller $\widehat C_3$, reflecting the stronger temporal dependence and retained nonlinear processing required by higher-order NARMA tasks.

Taken together, all benchmark tasks exhibit optimized hyperparameter settings within the active regions identified in the preceding IPC analysis, consistent with their respective temporal and nonlinear processing characteristics. These observations highlight the importance of tailored QRC hyperparameter optimization.

\section{Real-world prediction tasks}
\label{sec:prediction-tasks}

To evaluate the proposed QRC on real-world data, we apply it to prediction tasks based on the six real-world time-series datasets considered in Ref.~\citep{li2023multireservoir}. As baselines, we select three widely used multi-reservoir echo state networks. In Subsec.~\ref{subsec:prediction-data-baselines}, we introduce the six datasets and the MRESNs used for comparison. In Subsec.~\ref{subsec:prediction-results}, we present the prediction results. The proposed QRC achieves performance that is at least competitive with the MRESN baselines in most cases and outperforms all three on over half of the tasks.

\subsection{Datasets and baseline models}
\label{subsec:prediction-data-baselines}

We consider the following six real-world time-series datasets.
\begin{description}
  \item[Cardio] A dataset containing daily numbers of cardiovascular inpatients in Hong Kong hospitals.
  \item[Sunspot] A dataset containing monthly sunspot numbers from January, 1749 to November, 2019.
  \item[DMTMA] A dataset containing daily maximum temperatures measured at Melbourne airport from January 1, 2006 to December 31, 2019.
  \item[Electricity] A dataset containing daily electricity consumption of open power system data in Germany from January 1, 2006 to December 31, 2017.
  \item[Bike Sharing] A dataset containing hourly bike-rental counts from 21:00, May 9, 2011 to 0:00, December 5, 2011.
  \item[Traffic Speed]  A dataset of urban traffic speeds recorded at 10-minute intervals on road segment No.~61 in Guangzhou from August 1 to September 30, 2016.
\end{description}
Representative time-series plots of the six real-world datasets are shown in Fig.~\ref{fig:prediction-datasets}. For each dataset, we consider $K$-step-ahead prediction, where $K\in\mathbb N$ is the prediction horizon. The target at time step $n$ is defined as $d[n]=u[n+K]$. We also consider multi-output prediction, in which multiple prediction horizons $K_1,\ldots,K_{N_{\mathrm{y}}}$ are predicted simultaneously. The corresponding target vector is $\mathbf d[n]={\left(u[n+K_1],\ldots, u[n+K_{N_{\mathrm{y}}}]\right)}$.

To assess the prediction performance of the proposed QRC, we compare our proposal with MRESN baselines. An echo state network (ESN) is a classical RC system in which a fixed recurrent neural network serves as the reservoir and only a linear readout is trained~\citep{jaeger2001echo}. An MRESN extends this architecture by combining multiple reservoirs to construct richer reservoir-state representations. Specifically, we consider the following three commonly used MRESNs. Further details of these baselines can be found in Ref.~\citep{li2023multireservoir}.

\begin{description}
  \item[DeepESN] The deep echo state network (DeepESN) stacks multiple reservoir layers through inter-reservoir connections \cite{gallicchio2017deep}. The successive layers provide a hierarchical dynamical representation of the input sequence. The outputs of all reservoir layers are collectively treated as the reservoir state and used by a single final readout.
  \item[MSLESN] The multi-step learning echo state network (MSLESN) connects multiple ESNs hierarchically \cite{akiyama2022computational}. Unlike DeepESN, however, its output weight matrices are learned sequentially, one reservoir at a time, rather than being trained jointly in a single step.
  \item[GroupedESN] The grouped echo state network (GroupedESN) arranges multiple independent reservoirs in parallel \cite{gallicchio2017deep}. Each reservoir receives the same input sequence, and their reservoir states are combined and used by a single final readout.
\end{description}

\subsection{Results}
\label{subsec:prediction-results}

We first apply min-max rescaling for each dataset to map the sequence values to the interval $[0,1]$. The lengths of the washout, training, validation, and test segments for the six datasets are summarized in Table~\ref{tab:real-world-data-partitions}. The ridge regularization strength is set to $\lambda=10^{-6}$. Each experiment is repeated over $20$ independent trials. We consider 4- and 6-qubit QRCs, whose feature dimensions $N_{\mathrm{x}}=4^{N_{\mathrm{q}}}$ are $256$ and $4096$, respectively. For comparison, the total number of reservoir nodes is at most $5500$ for each of DeepESN, MSLESN, and GroupedESN.

\begin{table}[!tbp]
  \centering
  \caption{Data partitions for the six real-world datasets.}%
  \label{tab:real-world-data-partitions}
  \tablelayout{1}
  \fittable{%
    \begin{tabular}{@{}lrrrr@{}}
      \toprule
      Dataset & Washout & Training & Validation & Test \\
      \midrule
      Cardio        & 131 & 500  & 200  & 200  \\
      Sunspot       & 250 & 2000 & 500  & 500  \\
      DMTMA         & 112 & 3000 & 1000 & 1000 \\
      Electricity   & 182 & 2200 & 1000 & 1000 \\
      Bike Sharing  & 100 & 2900 & 1000 & 1000 \\
      Traffic Speed & 783 & 4000 & 2000 & 2000 \\
      \bottomrule
    \end{tabular}%
  }
\end{table}

\begin{table*}[!tp]
  \centering
  \caption{NRMSE results (mean $\pm$ standard deviation) of the 4- and 6-qubit QRCs and the three baselines across the six real-world datasets for prediction horizon at $K\in\{1,3,5\}$ and multi-output prediction at all three horizons simultaneously.}
  \tablelayout[1pt]{1.08}
  \fittable{%
    \begin{tabular}{llcccc}
      \toprule
      Dataset & Model & $K=1$ & $K=3$ & $K=5$ & Multi-output \\
      \midrule
      Cardio
        & \ourscell{QRC (4 qubits)}
        & \ourscell{\resultcell{6.36E-01}{1.25E-02}}
        & \ourscell{\resultcell{7.51E-01}{9.81E-03}}
        & \ourscell{\resultcell{7.59E-01}{7.08E-03}}
        & \ourscell{\resultcell{7.25E-01}{1.43E-02}} \\
      & \ourscell{QRC (6 qubits)}
        & \ourscell{\bestresultcell{6.31E-01}{7.16E-03}}
        & \ourscell{\resultcell{7.52E-01}{7.61E-03}}
        & \ourscell{\resultcell{7.50E-01}{7.65E-03}}
        & \ourscell{\bestresultcell{7.09E-01}{6.07E-03}} \\
      \cmidrule{2-6}
      & DeepESN & \resultcell{6.47E-01}{5.04E-03} & \resultcell{7.28E-01}{7.80E-03} & \resultcell{7.33E-01}{1.16E-02} & \resultcell{7.30E-01}{2.62E-03} \\
      & MSLESN & \resultcell{6.41E-01}{3.63E-03} & \bestresultcell{7.20E-01}{1.51E-02} & \bestresultcell{7.27E-01}{8.98E-03} & \resultcell{7.18E-01}{3.32E-03} \\
      & GroupedESN & \resultcell{6.40E-01}{4.33E-03} & \resultcell{7.27E-01}{5.04E-03} & \resultcell{7.36E-01}{1.23E-02} & \resultcell{7.27E-01}{3.99E-03} \\
      \midrule
      Sunspot
        & \ourscell{QRC (4 qubits)}
        & \ourscell{\resultcell{3.20E-01}{3.00E-04}}
        & \ourscell{\resultcell{3.64E-01}{1.88E-03}}
        & \ourscell{\bestresultcell{3.72E-01}{2.07E-03}}
        & \ourscell{\resultcell{3.55E-01}{1.63E-03}} \\
      & \ourscell{QRC (6 qubits)}
        & \ourscell{\resultcell{3.21E-01}{4.80E-04}}
        & \ourscell{\bestresultcell{3.59E-01}{9.20E-04}}
        & \ourscell{\bestresultcell{3.72E-01}{1.40E-03}}
        & \ourscell{\bestresultcell{3.52E-01}{7.05E-04}} \\
      \cmidrule{2-6}
      & DeepESN & \resultcell{3.19E-01}{6.56E-04} & \resultcell{3.64E-01}{1.32E-03} & \resultcell{3.82E-01}{1.96E-03} & \resultcell{3.55E-01}{1.79E-03} \\
      & MSLESN & \bestresultcell{3.16E-01}{8.65E-04} & \resultcell{3.63E-01}{1.56E-03} & \resultcell{3.83E-01}{3.29E-03} & \resultcell{3.57E-01}{1.93E-03} \\
      & GroupedESN & \resultcell{3.20E-01}{3.82E-04} & \resultcell{3.63E-01}{1.08E-03} & \resultcell{3.82E-01}{2.44E-03} & \bestresultcell{3.52E-01}{6.12E-04} \\
      \midrule
      DMTMA
        & \ourscell{QRC (4 qubits)}
        & \ourscell{\resultcell{5.97E-01}{5.18E-04}}
        & \ourscell{\resultcell{6.71E-01}{6.58E-04}}
        & \ourscell{\resultcell{6.75E-01}{4.46E-04}}
        & \ourscell{\resultcell{6.46E-01}{1.48E-03}} \\
      & \ourscell{QRC (6 qubits)}
        & \ourscell{\bestresultcell{5.87E-01}{5.62E-04}}
        & \ourscell{\bestresultcell{6.68E-01}{1.73E-03}}
        & \ourscell{\resultcell{6.69E-01}{1.21E-03}}
        & \ourscell{\bestresultcell{6.43E-01}{1.19E-03}} \\
      \cmidrule{2-6}
      & DeepESN & \resultcell{5.92E-01}{7.53E-04} & \bestresultcell{6.68E-01}{1.80E-03} & \resultcell{6.72E-01}{4.87E-04} & \resultcell{6.51E-01}{6.44E-04} \\
      & MSLESN & \resultcell{5.94E-01}{2.28E-03} & \resultcell{6.69E-01}{8.24E-04} & \bestresultcell{6.67E-01}{2.13E-03} & \resultcell{6.54E-01}{9.23E-04} \\
      & GroupedESN & \resultcell{5.94E-01}{2.28E-03} & \bestresultcell{6.68E-01}{1.80E-03} & \bestresultcell{6.67E-01}{2.13E-03} & \resultcell{6.52E-01}{2.58E-04} \\
      \midrule
      Electricity
        & \ourscell{QRC (4 qubits)}
        & \ourscell{\resultcell{2.94E-01}{1.77E-02}}
        & \ourscell{\resultcell{3.86E-01}{1.50E-02}}
        & \ourscell{\resultcell{4.29E-01}{1.19E-02}}
        & \ourscell{\resultcell{3.78E-01}{1.51E-02}} \\
      & \ourscell{QRC (6 qubits)}
        & \ourscell{\resultcell{2.79E-01}{5.19E-03}}
        & \ourscell{\resultcell{3.69E-01}{8.55E-03}}
        & \ourscell{\bestresultcell{4.09E-01}{9.61E-03}}
        & \ourscell{\bestresultcell{3.56E-01}{7.01E-03}} \\
      \cmidrule{2-6}
      & DeepESN & \resultcell{2.79E-01}{3.80E-03} & \resultcell{3.70E-01}{3.30E-03} & \resultcell{4.11E-01}{3.48E-03} & \resultcell{3.65E-01}{2.91E-03} \\
      & MSLESN & \bestresultcell{2.68E-01}{3.00E-03} & \bestresultcell{3.60E-01}{6.11E-03} & \resultcell{4.11E-01}{4.61E-03} & \resultcell{3.64E-01}{3.65E-03} \\
      & GroupedESN & \resultcell{2.90E-01}{2.50E-02} & \resultcell{3.67E-01}{5.01E-03} & \resultcell{4.14E-01}{4.40E-03} & \resultcell{3.58E-01}{2.62E-03} \\
      \midrule
      Bike Sharing
        & \ourscell{QRC (4 qubits)}
        & \ourscell{\resultcell{3.77E-01}{1.68E-02}}
        & \ourscell{\resultcell{5.83E-01}{2.43E-02}}
        & \ourscell{\resultcell{6.26E-01}{1.29E-02}}
        & \ourscell{\resultcell{5.45E-01}{1.60E-02}} \\
      & \ourscell{QRC (6 qubits)}
        & \ourscell{\resultcell{3.11E-01}{1.36E-02}}
        & \ourscell{\bestresultcell{5.24E-01}{2.50E-02}}
        & \ourscell{\bestresultcell{6.09E-01}{4.00E-02}}
        & \ourscell{\resultcell{4.99E-01}{2.94E-02}} \\
      \cmidrule{2-6}
      & DeepESN & \resultcell{3.02E-01}{9.10E-03} & \resultcell{5.43E-01}{2.11E-02} & \resultcell{6.30E-01}{3.15E-02} & \resultcell{5.19E-01}{1.61E-02} \\
      & MSLESN & \bestresultcell{2.68E-01}{4.91E-03} & \resultcell{5.29E-01}{1.95E-02} & \resultcell{7.06E-01}{1.04E-02} & \bestresultcell{4.97E-01}{1.76E-02} \\
      & GroupedESN & \resultcell{3.03E-01}{6.42E-03} & \resultcell{5.29E-01}{1.95E-02} & \resultcell{6.34E-01}{3.95E-02} & \resultcell{5.23E-01}{1.63E-02} \\
      \midrule
      Traffic Speed
        & \ourscell{QRC (4 qubits)}
        & \ourscell{\resultcell{2.40E-01}{2.45E-03}}
        & \ourscell{\resultcell{3.88E-01}{3.64E-03}}
        & \ourscell{\resultcell{4.52E-01}{1.65E-02}}
        & \ourscell{\resultcell{3.82E-01}{3.17E-03}} \\
      & \ourscell{QRC (6 qubits)}
        & \ourscell{\resultcell{2.39E-01}{3.25E-03}}
        & \ourscell{\resultcell{3.52E-01}{3.30E-03}}
        & \ourscell{\resultcell{4.06E-01}{4.17E-03}}
        & \ourscell{\bestresultcell{3.35E-01}{2.58E-03}} \\
      \cmidrule{2-6}
      & DeepESN & \bestresultcell{2.31E-01}{2.77E-04} & \bestresultcell{3.43E-01}{2.16E-03} & \bestresultcell{4.02E-01}{5.06E-03} & \resultcell{3.57E-01}{3.02E-03} \\
      & MSLESN & \resultcell{2.32E-01}{5.58E-04} & \resultcell{3.51E-01}{2.39E-04} & \resultcell{4.38E-01}{7.10E-03} & \resultcell{3.53E-01}{1.50E-03} \\
      & GroupedESN & \bestresultcell{2.31E-01}{6.92E-04} & \resultcell{3.55E-01}{4.42E-03} & \resultcell{4.26E-01}{7.67E-03} & \resultcell{3.56E-01}{1.20E-03} \\
      \bottomrule
    \end{tabular}%
  }
  \label{tab:results-all}
\end{table*}

The NRMSE results (mean $\pm$ standard deviation) of the 4- and 6-qubit configurations of the proposed QRC and the three MRESN baselines on the 24 prediction tasks based on the six real-world datasets are reported in Table~\ref{tab:results-all}. The 4- and 6-qubit configurations outperform all three MRESNs on 4 and 13 tasks, respectively. On the remaining tasks, the 4-qubit configuration remains competitive in some cases, while the 6-qubit configuration generally remains competitive compared to the baselines. Overall, the 6-qubit configuration outperforms its 4-qubit counterpart.

\begin{table*}[!tp]
  \centering
  \caption{Optimized hyperparameter settings for the 4- and 6-qubit QRCs across the six real-world datasets. Each entry is formatted as $\alpha_{\mathrm{in}}/J/h_{\mathrm{z}}/\gamma_{\mathrm{ad}}/\gamma_{\mathrm{pd}}$. }
  \tablelayout[4pt]{1.18}
  \fittable{%
    \begin{tabular}{lcc@{\hspace{0.5cm}}cc}
      \toprule
      & \multicolumn{2}{c}{$K=1$} & \multicolumn{2}{c}{$K=3$} \\
      \cmidrule(lr){2-3}\cmidrule(lr){4-5}
      Dataset & QRC (4 qubits) & QRC (6 qubits) & QRC (4 qubits) & QRC (6 qubits) \\
      \midrule
      Cardio & \paramcell{0.11}{0.59}{0.079}{0.0046}{0.12} & \paramcell{0.098}{0.60}{0.095}{0.0056}{0.15} & \paramcell{0.078}{0.60}{0.12}{0.0046}{0.13} & \paramcell{0.12}{0.38}{0.15}{0.0022}{0.10} \\
      \addlinespace[2pt]
      Sunspot & \paramcell{0.48}{0.11}{0.18}{0.071}{0.56} & \paramcell{0.60}{0.11}{0.20}{0.039}{0.84} & \paramcell{0.46}{0.083}{0.23}{0.036}{0.50} & \paramcell{0.56}{0.11}{0.19}{0.035}{0.68} \\
      \addlinespace[2pt]
      DMTMA & \paramcell{0.83}{0.074}{2.4}{0.047}{3.1} & \paramcell{0.69}{0.19}{7.0}{0.010}{3.0} & \paramcell{0.90}{0.094}{8.8}{0.013}{3.7} & \paramcell{0.73}{0.26}{6.4}{0.014}{4.5} \\
      \addlinespace[2pt]
      Electricity & \paramcell{0.19}{0.53}{0.0}{0.10}{0.049} & \paramcell{0.16}{0.58}{0.51}{0.13}{0.06} & \paramcell{0.23}{0.58}{0.0}{0.097}{0.038} & \paramcell{0.14}{0.50}{0.0}{0.10}{0.12} \\
      \addlinespace[2pt]
      Bike Sharing & \paramcell{0.20}{0.34}{0.085}{0.012}{0.085} & \paramcell{0.23}{0.28}{0.050}{0.26}{0.029} & \paramcell{0.12}{0.33}{0.060}{0.0090}{0.035} & \paramcell{0.18}{0.34}{0.060}{0.25}{0.0050} \\
      \addlinespace[2pt]
      Traffic Speed
        & \paramcell{0.47}{2.1}{0.0}{0.0053}{0.20} & \paramcell{0.11}{0.51}{0.10}{0.000089}{0.0090} & \paramcell{0.47}{1.8}{0.0}{0.0024}{0.14} & \paramcell{0.40}{0.60}{0.030}{0.000030}{0.0050} \\
      \midrule
      & \multicolumn{2}{c}{$K=5$} & \multicolumn{2}{c}{Multi-output} \\
      \cmidrule(lr){2-3}\cmidrule(lr){4-5}
      Dataset & QRC (4 qubits) & QRC (6 qubits) & QRC (4 qubits) & QRC (6 qubits) \\
      \midrule
      Cardio & \paramcell{0.080}{0.58}{0.11}{0.0047}{0.17} & \paramcell{0.11}{0.36}{0.15}{0.0025}{0.13} & \paramcell{0.078}{0.60}{0.093}{0.0048}{0.13} & \paramcell{0.12}{0.38}{0.16}{0.0024}{0.11} \\
      \addlinespace[2pt]
      Sunspot & \paramcell{0.44}{0.071}{0.22}{0.036}{0.47} & \paramcell{0.55}{0.14}{0.20}{0.034}{0.64} & \paramcell{0.45}{0.083}{0.22}{0.037}{0.48} & \paramcell{0.56}{0.12}{0.19}{0.035}{0.66} \\
      \addlinespace[2pt]
      DMTMA & \paramcell{0.90}{0.015}{8.8}{0.012}{6.0} & \paramcell{0.71}{0.18}{7.2}{0.015}{4.3} & \paramcell{0.45}{0.071}{2.6}{0.013}{1.6} & \paramcell{0.75}{0.22}{6.9}{0.013}{3.3} \\
      \addlinespace[2pt]
      Electricity & \paramcell{0.26}{0.60}{0.010}{0.13}{0.071} & \paramcell{0.30}{1.1}{0.10}{0.00030}{0.015} & \paramcell{0.20}{0.58}{0.0}{0.090}{0.029} & \paramcell{0.19}{1.0}{0.10}{0.00070}{0.019} \\
      \addlinespace[2pt]
      Bike Sharing & \paramcell{0.10}{0.33}{0.060}{0.0090}{0.035} & \paramcell{0.17}{0.34}{0.060}{0.27}{0.0080} & \paramcell{0.13}{0.34}{0.060}{0.010}{0.046} & \paramcell{0.18}{0.34}{0.060}{0.26}{0.0070} \\
      \addlinespace[2pt]
      Traffic Speed & \paramcell{0.33}{3.8}{0.010}{0.0017}{0.026} & \paramcell{0.33}{0.65}{0.090}{0.000050}{0.0062} & \paramcell{0.46}{3.6}{0.040}{0.0018}{0.12} & \paramcell{0.33}{0.65}{0.090}{0.000070}{0.0068} \\
      \bottomrule
    \end{tabular}%
  }
  \label{tab:param-all}
\end{table*}

The Cardio and DMTMA datasets are relatively challenging for all evaluated models, as indicated by their results. For Cardio, both QRC configurations outperform all three MRESNs on the $K=1$ task; moreover, the 6-qubit configuration outperforms all three baselines on the multi-output task. For DMTMA, both configurations outperform all three MRESNs on the multi-output task; moreover, the 6-qubit configuration outperforms all three baselines on the $K=1$ and $K=3$ tasks. These results indicate that the proposed QRC remains competitive on relatively challenging datasets, with its advantage over the baselines becoming more pronounced as the number of qubits increases.

For Electricity and Traffic Speed datasets, the 4-qubit configuration is less competitive with the best-performing MRESNs than on the other datasets. However, the 6-qubit configuration achieves competitive performance on both datasets, outperforming all three MRESNs on the $K=5$ and multi-output Electricity tasks and on the multi-output Traffic Speed task. These results suggest that, in some cases, increasing the number of qubits can make the proposed QRC more competitive with classical baselines.

Finally, for Sunspot and Bike Sharing, both configurations outperform all three MRESNs on the $K=5$ tasks, while the 6-qubit configurations also outperform them on the $K=3$ and multi-output Sunspot tasks and on the $K=3$ Bike Sharing task.

The optimized hyperparameter settings for the 4- and 6-qubit QRC configurations across the 24 real-world prediction tasks are reported in Table~\ref{tab:param-all}. All tasks adopt dual-decoherence settings, while different prediction horizons of the same dataset generally correspond to distinct optimized settings within the same operating region. This pattern indicates similar information-processing requirements within each dataset, while the variation in the optimized decoherence rates reflects differences among the prediction horizons. Moreover, in most cases, the 4- and 6-qubit configurations share the same operating region. These results highlight the benefit of jointly tuning amplitude damping and pure dephasing for real-world prediction tasks.

\section{Conclusion}
\label{sec:discussion-conclusion}
We have proposed a QRC model that harnesses pure dephasing within a minimum complexity architecture. A fixed bipolar input weight matrix accommodates scalar and multidimensional inputs through Hamiltonian encoding, while an open uniform XX chain combines competitive computational performance with a structurally simple and deterministic design that may facilitate physical implementation. Building on strict contractivity as a sufficient condition for the ESP and the FMP, we show that no mixing time is required for QRC induced by relaxing QMSs with faithful steady states. Furthermore, under suitable input encoding, pure dephasing can ensure the ESP and the FMP for general XYZ reservoirs, including the proposed XX chain, even without amplitude damping. In the presence of amplitude damping, these reservoirs have distinct stationary states for distinct inputs. Introducing pure dephasing alters these states under amplitude damping alone and hence affects the reservoir dynamics. These results provide a theoretical basis for using pure dephasing as a computational resource for QRC.

IPC analysis reveals partially independent, non-additive, and degree-dependent responses of linear and nonlinear memory capacities to amplitude damping and pure dephasing. Joint control of the two decoherence rates allows the relative magnitudes of these capacities to be regulated, providing access to capacity profiles that may not be obtained by tuning either rate alone. These findings highlight the benefit of fine-tuning both decoherence rates for individual tasks.

The proposed QRC achieves promising performance across the five synthetic benchmark tasks. On the real-world prediction tasks, the 6-qubit configuration matches or outperforms the best-performing MRESN baseline on 13 of the 24 tasks, based on the reported mean NRMSE. These results suggest that the proposed quantum reservoir can generate sufficiently rich task-relevant information for nonlinear time-series prediction, even with only a few qubits. All optimized real-world configurations adopt dual-decoherence settings, highlighting the benefit of jointly tuning amplitude damping and pure dephasing.

\begin{acknowledgments}
This work was partly supported by JST CREST Grant Number JPMJCR24R2 (Y.S. and G.T.), JSPS KAKENHI Grant Numbers JP23K28154 and JP25K24744 (G.T.), NSFC Grant Numbers U25A6009 and 92565301 (H.F.). We thank Tomohiro Hayashi of Nagoya Institute of Technology for valuable assistance on the proof of Theorem~\ref{thm:part2-positive-contractive}. We also thank Ziqiang Li of Nagoya Institute of Technology for providing information on the synthetic benchmark and real-world prediction tasks, and Kaixuan Huang of Beijing Academy of Quantum Information Sciences for fruitful discussion.

\end{acknowledgments}

\section*{Data availability}
The data and code supporting the findings of this work are available from the contact authors upon request during peer review. These materials will be made publicly available upon publication.

\appendix

\section{Proofs}
\label{app:proofs}

\subsection{Proof of Theorem~\ref{thm:part2-positive-contractive}}
\label{app:strict-positivity-contractivity}

\begin{proof}
Let $d=\dim\mathcal X$. It suffices to consider $d\ge2$. Define $\mathcal A:=\{X\in\Lin(\mathcal X):X=X^\ast,\ \operatorname{Tr}(X)=0,\ \|X\|_1=1\}$. For a fixed $X\in\mathcal A$, let $X=X_+-X_-$ be its decomposition into positive and negative parts, so that $X_+,X_-\succeq0$ and $X_+X_-=0$. Since $\operatorname{Tr}(X)=0$ and $\|X\|_1=1$, it follows that $\operatorname{Tr}(X_+)=\operatorname{Tr}(X_-)=1/2$. In particular, both $X_+$ and $X_-$ are nonzero.

For any self-adjoint operators $P$ and $Q$, let $m(P,Q)$ denote the maximum of $\operatorname{Tr}(Y)$ over all self-adjoint operators $Y$ satisfying $Y\preceq P$ and $Y\preceq Q$. We recall the following identity,
\begin{equation}
  m(P,Q)
  =\frac{1}{2}\operatorname{Tr}(P+Q)
  -\frac{1}{2}\|P-Q\|_1.
  \label{eq:common-lower-bound-trace}
\end{equation}
Set $P=\Phi(X_+)$ and $Q=\Phi(X_-)$. Strict positivity of $\Phi$ implies that $P$ and $Q$ are positive definite. Hence, for $\varepsilon=\min\{\lambda_{\min}(P),\lambda_{\min}(Q)\}>0$, the operator $\varepsilon I$ satisfies $\varepsilon I\preceq P$ and $\varepsilon I\preceq Q$. Consequently, $m(P,Q)\ge d\varepsilon>0$.

Since $\Phi$ is trace preserving and linear, $\operatorname{Tr}(P+Q)=1$ and $P-Q=\Phi(X)$. Substituting these identities into Eq.~\eqref{eq:common-lower-bound-trace} yields $0<m(P,Q)=\frac{1}{2}\left(1-\|\Phi(X)\|_1\right)$. Therefore $\|\Phi(X)\|_1<1$ for every $X\in\mathcal A$. By compactness of $\mathcal A$ and continuity of $X\mapsto\|\Phi(X)\|_1$, there exists $X_0\in\mathcal A$ such that $c=\max_{X\in\mathcal A}\|\Phi(X)\|_1 =\|\Phi(X_0)\|_1<1$.

Finally, let $\rho,\sigma\in\Den(\mathcal X)$. If $\rho=\sigma$, the desired inequality holds trivially. Suppose that $\rho\neq\sigma$, and set $X=(\rho-\sigma)/\|\rho-\sigma\|_1$. Then $X\in\mathcal A$. By linearity of $\Phi$, we obtain $\|\Phi(\rho)-\Phi(\sigma)\|_1 \le c\|\rho-\sigma\|_1$.
\end{proof}

\subsection{Proof of Proposition~\ref{prop:part2-uniform-primitivity-index}}
\label{app:uniform-primitivity-index}

\begin{proof}
Fix $\mathbf u\in D_{N_{\mathrm{u}}}$ and choose $k_{\mathbf u}$ such that ${T(\cdot,\mathbf u)}^{k_{\mathbf u}}$ is strictly positive. For each $k\in\mathbb N$, define $f_k(\mathbf v):=\min_{\rho\in\Den(\mathcal X)} \lambda_{\min}\!\left({T(\cdot,\mathbf v)}^k(\rho)\right)$. The compactness of $\Den(\mathcal X)$ and the continuity of $T$ imply that $f_k$ is continuous. Moreover, ${T(\cdot,\mathbf v)}^k$ is strictly positive if and only if $f_k(\mathbf v)>0$. Since $f_{k_{\mathbf u}}(\mathbf u)>0$, there exists a neighborhood $U_{\mathbf u}$ of $\mathbf u$ such that $f_{k_{\mathbf u}}(\mathbf v)>0$ for every $\mathbf v\in U_{\mathbf u}$. Hence ${T(\cdot,\mathbf v)}^{k_{\mathbf u}}$ is strictly positive for every $\mathbf v\in U_{\mathbf u}$. Compactness of $D_{N_{\mathrm{u}}}$ gives a finite subcover $U_{\mathbf u_1},\ldots,U_{\mathbf u_s}$. Set $W(T):=\max_{1\le j\le s}k_{\mathbf u_j}$. If $\mathbf v\in U_{\mathbf u_j}$ and $A\ge0$ is nonzero, then ${T(\cdot,\mathbf v)}^{W(T)-k_{\mathbf u_j}}(A)$ remains a nonzero positive operator because the channel is trace preserving. Applying the strictly positive map ${T(\cdot,\mathbf v)}^{k_{\mathbf u_j}}$ therefore gives ${T(\cdot,\mathbf v)}^{W(T)}(A)>0$.
\end{proof}

\subsection{Proof of Proposition~\ref{prop:part2-proposed-reservoir-algebra}}
\label{app:proposed-reservoir-algebra}

\begin{proof}
Since $\gamma^{\mathrm{pd}}_j>0$ for every $j\in V$, each $Z_j=\left(\gamma^{\mathrm{pd}}_j\right)^{-1/2}L_j^{\mathrm{pd}}$ belongs to $\mathcal A$. The operator $D=\sum_\mu L_\mu^\ast L_\mu$ is a linear combination of the identity and the operators $Z_j$, and therefore $D\in\mathcal A$. Since $K=iH+D/2$, it follows that $H=-i(K-D/2)\in\mathcal A$.

For $S\subseteq\{1,\ldots,N\}$ and $\boldsymbol\epsilon\in{\{0,1\}}^N$, write $Z^{\boldsymbol\epsilon}:=\prod_{j=1}^{N}Z_j^{\epsilon_j}$ and $\chi_S(\boldsymbol\epsilon):={\left(-1\right)}^{ \sum_{j\in S}\epsilon_j}$. Define the projection $\Pi_S(B):=2^{-N}\sum_{\boldsymbol\epsilon\in{\{0,1\}}^N} \chi_S(\boldsymbol\epsilon)Z^{\boldsymbol\epsilon}B Z^{\boldsymbol\epsilon}$. Since $\mathcal A$ contains every $Z_j$, it contains $\Pi_S(B)$ whenever it contains $B$. The map $\Pi_S$ extracts the Pauli strings whose $X$-or-$Y$ support is exactly $S$. For Eq.~\eqref{eq:general-anisotropic-heisenberg}, the interaction terms have $X$-or-$Y$ support either empty or of cardinality two. Therefore, for every $j\in V$, $A_j:=\Pi_{\{j\}}(H)=h_j^x X_j+h_j^y Y_j$ belongs to $\mathcal A$. Since $Z_j\in\mathcal A$, the operator $B_j:=-i Z_j A_j=-h_j^y X_j+h_j^x Y_j$ also belongs to $\mathcal A$. The coefficient matrix relating $(A_j,B_j)$ to $(X_j,Y_j)$ has determinant ${\left(h_j^x\right)}^2+{\left(h_j^y\right)}^2>0$. Hence $X_j,Y_j\in\mathcal A$ for every $j\in V$. Thus $\mathcal A$ contains the local operators $I$, $X_j$, $Y_j$, and $Z_j$ at every site. Their products form the Pauli basis of $\Lin(\mathcal X)$, and hence $\mathcal A=\Lin(\mathcal X)$.
\end{proof}

\begin{figure*}[!tp]
  \centering
  \widefiguregraphics{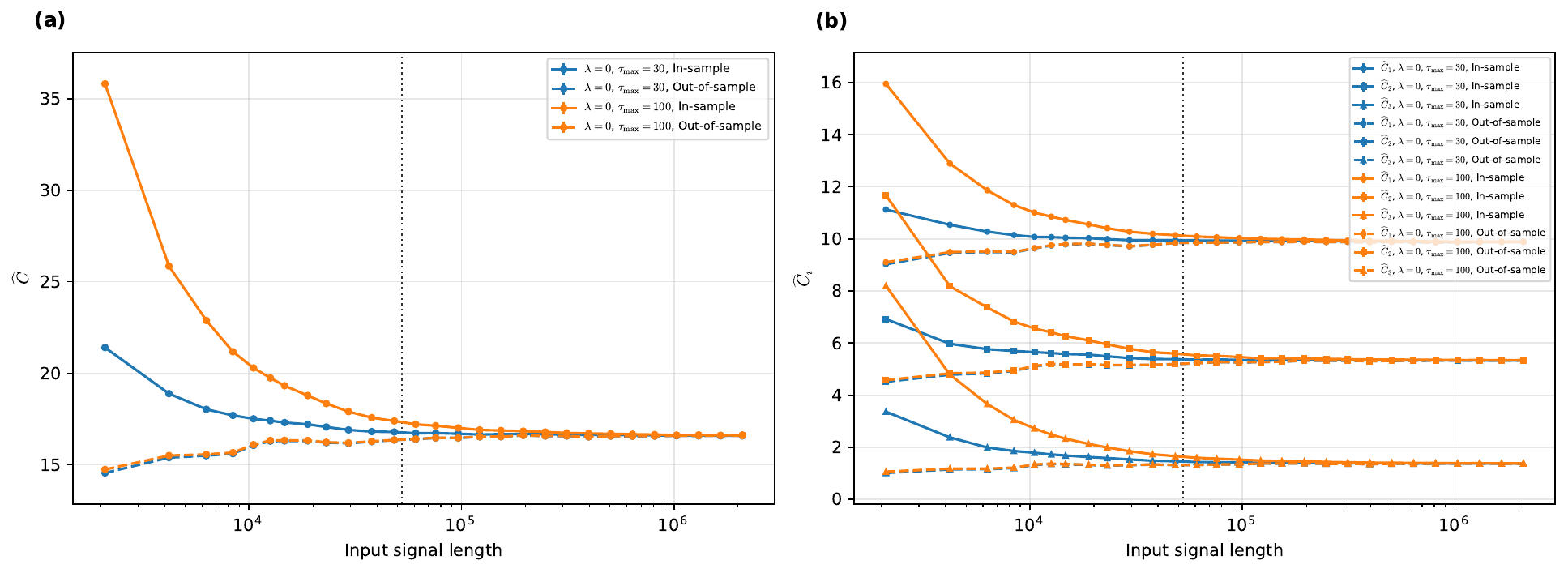}
  \caption{Dependence of (a) $\widehat{C}=\widehat C_1+\widehat C_2+\widehat C_3$ and (b) $\widehat{C}_i$ $(i=1,2,3)$ on signal length. In both panels, solid and dashed lines denote in-sample and out-of-sample settings, respectively, while blue and orange lines correspond to $(\lambda,\tau_{\max})=(0,30)$ and $(0,100)$, respectively. In panel~(b), circle, square, and triangle marks denote $\widehat{C}_1$, $\widehat{C}_2$, and $\widehat{C}_3$, respectively. Vertical dotted lines mark the selected input signal length of $52500$.}
  \label{fig:IPC_signal_length}
\end{figure*}

\subsection{Proof of Lemma~\ref{lem:hamiltonian-stationary-states}}
\label{app:proof-hamiltonian-stationary-states}

\begin{proof}
Let $L_1$ and $L_2$ denote the generators of $\Phi_1$ and $\Phi_2$. Then $( L_2- L_1)(\rho)=-i[\Delta H,\rho]$. If $\rho_1=\rho_2$, subtraction of the stationary equations gives $[\Delta H,\rho_1]=0$. Conversely, this commutation relation gives $ L_2(\rho_1)=0$, and uniqueness yields $\rho_1=\rho_2$. This proves \textnormal{(i)}.

Suppose that $\rho_1=\rho_2=\rho$. Then $[\Delta H,\rho]=[\Delta H,\rho^2]=0$ by \textnormal{(i)}. For each $j$, set $W_j=\Delta h_j^xY_j-\Delta h_j^yX_j$. The Pauli commutation relations and cyclicity of the trace give
\begin{align*}
 0
 &=\operatorname{Tr}(\rho^2[\Delta H,W_j])\\
 &=2i\bigl((\Delta h_j^x)^2+(\Delta h_j^y)^2\bigr)
   \operatorname{Tr}(\rho^2Z_j).
\end{align*}
The assumptions therefore imply $\operatorname{Tr}(\rho^2\sigma_j^z)=0$ for every $j$.

For any operator $A$, the Lindblad dissipator $D[A](\rho)=A\rho A^\ast-\tfrac12\{A^\ast A,\rho\}$ satisfies
\[
 \operatorname{Tr}(\rho D[A](\rho))
 =-\frac12\|[A,\rho]\|_2^2
  +\frac12\operatorname{Tr}(\rho^2[A,A^\ast]),
\]
where $\|A\|_2^2=\operatorname{Tr}(A^\ast A)$. Using $[\sigma_j^-,\sigma_j^+]=-Z_j$ and the stationary equation for $\rho$, we obtain
\begin{align*}
 0=\operatorname{Tr}(\rho L_1(\rho))
 ={}&-\frac12\sum_j\gamma_j^{\mathrm{ad}}
       \|[\sigma_j^-,\rho]\|_2^2\\
    &-\frac12\sum_j\gamma_j^{\mathrm{pd}}
       \|[Z_j,\rho]\|_2^2.
\end{align*}
Since every $\gamma_j^{\mathrm{ad}}>0$, it follows that $[\sigma_j^-,\rho]=0$ for every $j$. Hermiticity of $\rho$ also gives $[\sigma_j^+,\rho]=0$. The local lowering and raising operators generate the full operator algebra, so $\rho=I_{2^N}/2^N$. However, $ L_1(I_{2^N}/2^N)=-2^{-N}\sum_j\gamma_j^{\mathrm{ad}}Z_j\neq0$, a contradiction. This proves \textnormal{(ii)}.
\end{proof}

\subsection{Proof of Lemma~\ref{lem:ad-pd-stationary-state}}
\label{app:proof-ad-pd-stationary-state}

\begin{proof}
Let $L_1$ and $L_2$ denote the generators of $\Phi_1$ and $\Phi_2$. If $\rho_1=\rho_2$, their stationary equations give $( L_2-L_1)(\rho_1)=0$. Taking the Hilbert--Schmidt inner product with $\rho_1$ yields
\[
 0
 =\operatorname{Tr}\!\left[
      \rho_1( L_2- L_1)(\rho_1)\right]\nonumber
 =-\frac12\sum_{j=1}^{N}\gamma_j^{\mathrm{pd}}
       \bigl\|[\rho_1,\sigma_j^z]\bigr\|_2^2.
\]
It follows that $[\rho_1,\sigma_j^z]=0$ whenever $\gamma_j^{\mathrm{pd}}>0$. Conversely, these commutation relations imply $( L_2- L_1)(\rho_1)=0$ and hence $ L_2(\rho_1)=0$. The uniqueness of $\rho_2$ then gives $\rho_1=\rho_2$, proving \textnormal{(i)}.

Suppose that $\gamma_j^{\mathrm{pd}}>0$ for every $j$ and $\rho_1=\rho_2$. By the first statement, $\rho_1$ is diagonal in the joint eigenbasis of $\{Z_j\}_{j=1}^{N}$. Define $ N=\sum_{j=1}^{N}\sigma_j^+\sigma_j^-$. Since $[\rho_1, N]=0$, the stationary equation for $\Phi_1$ implies
\begin{equation}
\label{eq:ad-stationary-number-balance}
 0=\operatorname{Tr}[N L_1(\rho_1)]
  =-\sum_{j=1}^{N}\gamma_j^{\mathrm{ad}}
       \operatorname{Tr}(\sigma_j^+\sigma_j^-\rho_1).
\end{equation}
Since $\rho_1$ and $\sigma_j^+\sigma_j^-$ are diagonal with nonnegative entries and every $\gamma_j^{\mathrm{ad}}>0$, Eq.~\eqref{eq:ad-stationary-number-balance} implies $\sigma_j^+\sigma_j^-\rho_1=0$ for every $j$. Consequently, $\rho_0\rho_1=\rho_1$, since $\rho_0=\prod_{j=1}^{N}(I-\sigma_j^+\sigma_j^-)$. As $\rho_0$ has exactly one nonzero diagonal entry, normalization gives $\rho_1=\rho_0$. The stationary equation reduces to $[H,\rho_0]=0$. Conversely, $[H,\rho_0]=0$ makes $\rho_0$ stationary for both semigroups, and uniqueness yields $\rho_1=\rho_2=\rho_0$, proving \textnormal{(ii)}.

\begin{figure*}[!tp]
  \centering
  \widefiguregraphics{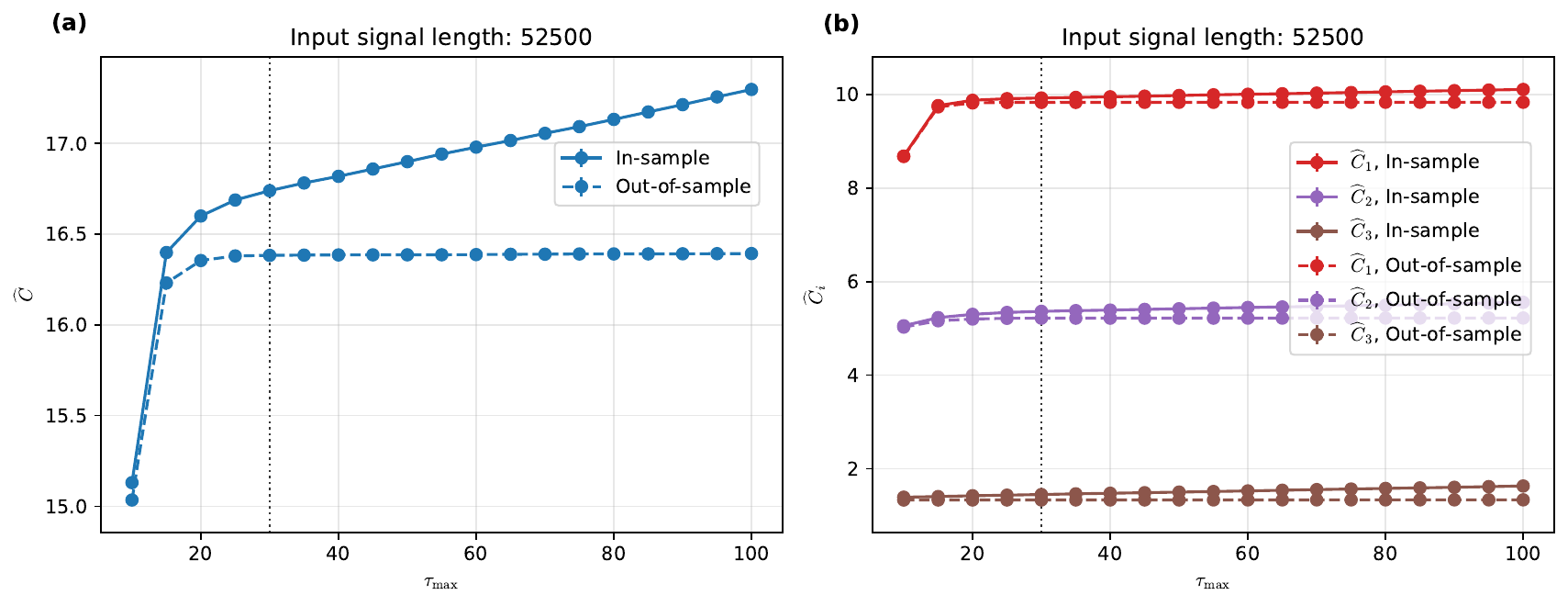}
  \caption{Dependence of (a) $\widehat C$ and (b) $\widehat{C}_i$ $(i=1,2,3)$ on the maximum delay $\tau_{\max}$ at the selected input signal length of $52500$ and $\lambda=0$. In both panels, solid and dashed lines denote in-sample and out-of-sample settings, respectively. In panel~(b), red, purple, and brown lines correspond to $\widehat{C}_1$, $\widehat{C}_2$, and $\widehat{C}_3$, respectively. Vertical dotted lines mark the selected maximum delay $\tau_{\max}=30$.}
  \label{fig:IPC_max_delay}
\end{figure*}

Finally, for the Hamiltonian in Eq.~\eqref{eq:general-anisotropic-heisenberg}, a direct calculation gives
\[
 \frac12\|[H,\rho_0]\|_2^2
 =\sum_{j=1}^{N}\left[(h_j^x)^2+(h_j^y)^2\right]
  +\sum_{\{j,k\}\in E}(J_{jk}^x-J_{jk}^y)^2.
\]
The right-hand side vanishes if and only if $h_j^x=h_j^y=0$ for all $j$ and $J_{jk}^x=J_{jk}^y$ for all $\{j,k\}\in E$. Together with \textnormal{(ii)}, this proves \textnormal{(iii)}.
\end{proof}

\subsection{Proof of Proposition~\ref{prop:ipc-properties}}
\label{app:ipc-properties}

\begin{proof}
Since an orthogonal projection is contractive, $0\le\|P_{\mathcal R}z\|_{\mathcal H}\le\|z\|_{\mathcal H}$, which proves the first inequality. Equality $\|P_{\mathcal R}z\|_{\mathcal H}=\|z\|_{\mathcal H}$ holds if and only if $z\in\mathcal R$, whereas $\|P_{\mathcal R}z\|_{\mathcal H}=0$ holds if and only if $z\in\mathcal R^\perp$. This proves \textnormal{(i)}.

Let ${\{e_j\}}_{j=1}^{r}$ be an orthonormal basis of $\mathcal R$. For every $\alpha\in\mathcal I$, Parseval's identity in $\mathcal R$ gives $\|P_{\mathcal R}z_\alpha\|_{\mathcal H}^{2} =\sum_{j=1}^{r}|\langle e_j,z_\alpha\rangle_{\mathcal H}|^2$. Summing over $\alpha$ and applying Parseval's identity in $\mathcal T$ yield
\begin{align}
  C_{\mathrm{tot}}(\mathcal R;\mathcal T)
  &=\sum_{\alpha\in\mathcal I}\sum_{j=1}^{r}
  |\langle e_j,z_\alpha\rangle_{\mathcal H}|^2 \notag\\
  &=\sum_{j=1}^{r}\|P_{\mathcal T}e_j\|_{\mathcal H}^{2}.
  \label{eq:ipc-total-projection-formula}
\end{align}
The right-hand side is independent of $\mathcal Z$, proving \textnormal{(ii)}.

Since $P_{\mathcal T}$ is contractive, Eq.~\eqref{eq:ipc-total-projection-formula} implies $C_{\mathrm{tot}}(\mathcal R;\mathcal T)\le r$. Equality holds if and only if $\|P_{\mathcal T}e_j\|_{\mathcal H}=1$ for every $j$, which is equivalent to $\mathcal R\subseteq\mathcal T$. If $\mathcal T$ is finite-dimensional, summing $C_{\mathcal R}(z_\alpha)\le1$ over an orthonormal basis of $\mathcal T$ also gives $C_{\mathrm{tot}}(\mathcal R;\mathcal T)\le\dim\mathcal T$. Taking $\mathcal T=\mathcal H$ proves the remaining assertions in \textnormal{(iii)}.

Finally, suppose that $\mathcal R_1\subseteq\mathcal R_2$. The orthogonal decomposition $P_{\mathcal R_2}z=P_{\mathcal R_1}z+ (P_{\mathcal R_2}-P_{\mathcal R_1})z$ gives $\|P_{\mathcal R_1}z\|_{\mathcal H}\le \|P_{\mathcal R_2}z\|_{\mathcal H}$. Hence $C_{\mathcal R_1}(z)\le C_{\mathcal R_2}(z)$. Summing this inequality over an orthonormal basis of $\mathcal T$ proves the corresponding inequality for the total IPC.
\end{proof}

\begin{figure*}[!tp]
  \centering
  \widefiguregraphics{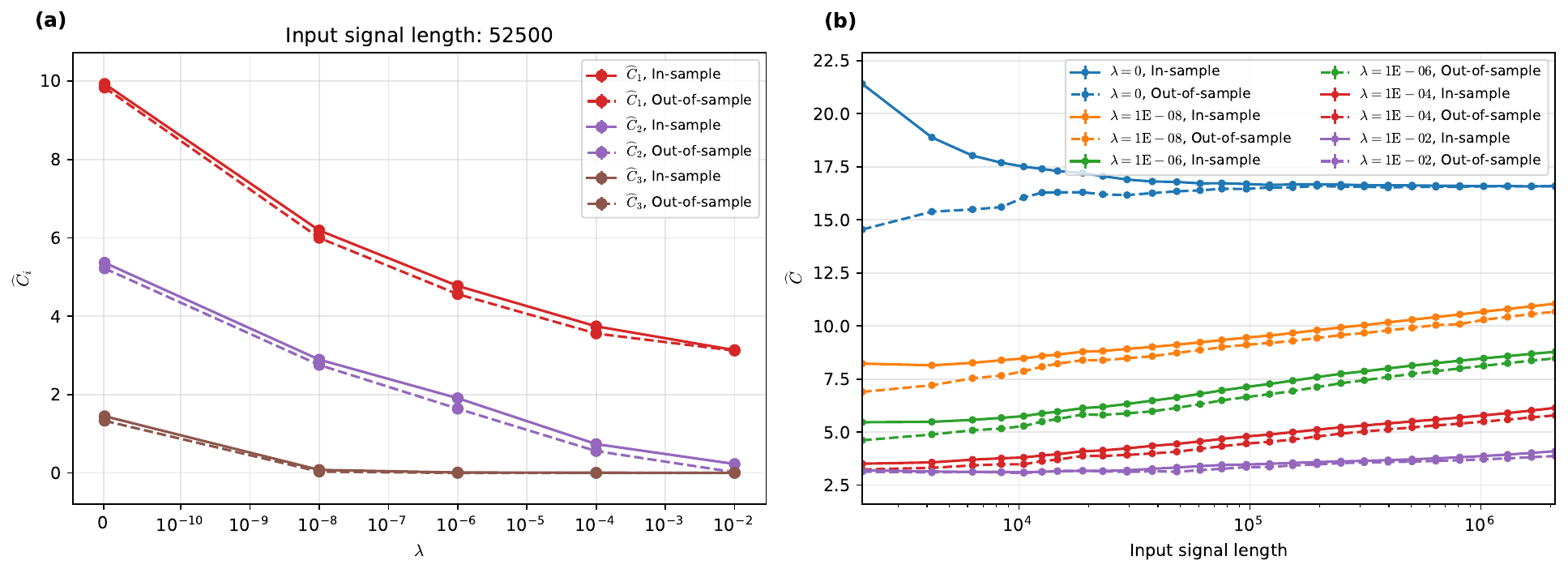}
  \caption{Effect of ridge regularization strength $\lambda$ on $\widehat C$ and $\widehat C_i$ $(i=1,2,3)$. Panel~(a) shows the dependence of $\widehat C_i$ on $\lambda$ at the selected input signal length of $52500$ and $\tau_{\max}=30$. Red, purple, and brown lines correspond to $\widehat{C}_1$, $\widehat{C}_2$, and $\widehat{C}_3$, respectively. Panel~(b) shows the dependence of $\widehat{C}$ on the input signal length at $\tau_{\max}=30$. Blue, orange, green, red, and purple lines correspond to $\lambda=0$, $10^{-8}$, $10^{-6}$, $10^{-4}$, and $10^{-2}$, respectively. In both panels, solid and dashed lines represent in-sample and out-of-sample settings, respectively.}
  \label{fig:IPC_ridge}
\end{figure*}

\section{Convergence analysis of empirical memory capacities}
\label{app:ipc-convergence}

For the 4-qubit QRC, we assess the numerical convergence of empirical memory capacities $\widehat C_i$ ($i=1,2,3$) with respect to the maximum delay $\tau_{\max}$ and the input signal length in both in-sample and out-of-sample settings. For the in-sample setting, each input sequence is divided into washout and training segments with a ratio of $1:20$, and the $\widehat C_i$ is computed on the training segment. Correspondingly, for the out-of-sample setting, the input sequence is divided into washout, training, and test segments with a ratio of $1:10:10$, where the $\widehat C_i$ is computed on the test segment. In addition, we also examine the dependence of empirical memory capacities on ridge regularization strength $\lambda$. Through these analyses, the suitable parameter setting is identified to balance approximation accuracy against computational cost.

Figure~\ref{fig:IPC_signal_length} shows the dependence of $\widehat C_i$, as well as $\widehat C=\widehat C_1+\widehat C_2+\widehat C_3$, on the input signal length. For a sufficiently large maximum delay, $\tau_{\max}=100$, with $\lambda=0$, the in-sample $\widehat{C}$ and $\widehat{C}_i$ exceed their out-of-sample counterparts; as the input signal length increases, the former gradually decrease while the latter increase, and the two become nearly indistinguishable when the input signal is sufficiently long. Reducing the maximum delay to $\tau_{\max}=30$ yields similar qualitative behavior, with $\widehat{C}$ and $\widehat{C}_i$ approaching the same converged values as those obtained for $\tau_{\max}=100$ as the input signal length increases. To reduce the computational cost, the out-of-sample results for $\tau_{\max}=30$ and $\lambda=0$ are used to determine an appropriate input signal length. We select an input signal length of $52500$, at which the relative deviations of $\widehat C$ and $\widehat C_i$ from their respective converged values are all below $5\%$.

Figure~\ref{fig:IPC_max_delay} shows the dependence of $\widehat C$ and $\widehat C_i$ on $\tau_{\max}$ at the selected input signal length of $52500$ and $\lambda=0$. Under the out-of-sample setting, both $\widehat C$ and $\widehat C_i$ rapidly approach their respective converged values as $\tau_{\max}$ increases. In contrast, their in-sample counterparts continue to increase throughout the range of $\tau_{\max}$ examined. Together with the results in Fig.~\ref{fig:IPC_signal_length}, this behavior indicates that longer input signals are required for the in-sample $\widehat C$ and $\widehat C_i$ to converge. For the out-of-sample setting with $\tau_{\max}=30$, the relative deviations of $\widehat C$ and $\widehat C_i$ from their respective converged values are below $1\%$. Therefore, $\tau_{\max}=30$ provides an appropriate balance between numerical convergence and computational cost.

Figure~\ref{fig:IPC_ridge} summarizes the effect of ridge regularization strength on $\widehat C$ and $\widehat C_i$. At the selected input signal length of $52500$ and $\tau_{\max}=30$, panel~(a) shows that $\widehat{C}_i$ decreases with increasing $\lambda$ under both the in-sample and out-of-sample settings. Panel~(b) further shows that, as the input signal length increases, the values of $\widehat{C}$ obtained with different $\lambda$ gradually approach the converged value for $\lambda=0$. These results indicate that larger values of $\lambda$ require longer input signals to achieve convergence. We therefore set $\lambda=0$ in the IPC analysis.

\bibliographystyle{quantum}
\bibliography{references}

\end{document}